\documentclass[review]{elsarticle}
\usepackage{lineno,hyperref}
\modulolinenumbers[5]

\journal{Elsevier}

\usepackage{tabularx}

\usepackage{graphicx}
\usepackage{latexsym}
\usepackage{color}
\usepackage{caption}
\usepackage{subcaption}
\usepackage{bm}
\usepackage{bbm}
\usepackage{amsmath,amssymb, amsthm}
\usepackage{algorithm}
\usepackage{algpseudocode}
\usepackage{color,soul}
\usepackage{graphics}
\usepackage{amsfonts}
\usepackage{setspace}   
\usepackage{wasysym}
\usepackage{multirow}

\usepackage{adjustbox}
\usepackage{graphicx}
\usepackage{multirow}
\usepackage{nomencl}
\usepackage{float}
\usepackage[table]{xcolor}
\makenomenclature
\newcommand*\diff{\mathop{}\!\mathrm{d}}
\usepackage{geometry}
\newtheorem{theorem}{Theorem}

\theoremstyle{definition}

\begin{document}

\title{A near-optimal sampling strategy for\\  sparse recovery~of polynomial~chaos~expansions}
\author{Negin Alemazkoor}
\author{Hadi Meidani}
\address{Department of Civil and Environmental Engineering, University of Illinois at Urbana-Champaign, Urbana, Illinois, USA.}

\begin{abstract}
Compressive sampling has  become  a  widely used approach to construct polynomial chaos  surrogates when the number of available simulation samples is limited. Originally, these expensive simulation samples would be obtained at random locations in the parameter space. It was later shown that the choice of sample locations could significantly impact the accuracy of  resulting surrogates. This motivated new sampling strategies or design-of-experiment approaches, such as coherence-optimal sampling, which aim at improving the coherence property. In this paper, we propose a sampling strategy that can identify near-optimal sample  locations that lead to improvement in  local-coherence property and also enhancement of cross-correlation properties of  measurement matrices.  We provide theoretical motivations for the proposed sampling strategy along with several numerical examples that show that our near-optimal sampling strategy produces substantially more  accurate results, compared to other sampling strategies.
\end{abstract}

\maketitle

\section{Introduction} \label{sec:intro}
In order to facilitate  stochastic computation in  analysis and design of complex systems, analytical surrogates that approximate and replace  full-scale simulation models have been increasingly studied. One of the most widely adopted surrogates is the polynomial chaos expansion (PCE), which approximates the quantity of interest (QoI) by a spectral representation using polynomial functions of random parameters \cite{ghanem2003stochastic, xiu2002wiener,deb2001solution, babuska2004galerkin}. In estimating these spectral surrogates, non-intrusive stochastic techniques, based on either spectral projection or linear regression, are widely used especially because they don't require modifying deterministic solvers or legacy codes, which is an otherwise  cumbersome task  \cite{eldred2008evaluation}. These non-intrusive techniques are still the subject of ongoing research as the number of required samples for accurate surrogate estimation rapidly grows with the number of random  parameters, even when efficient techniques such as sparse grid are used \cite{novak1997curse,ganapathysubramanian2007sparse, xiu2005high, nobile2008sparse}. 

More recently,  researchers have developed techniques, based on compressive sampling (CS), that are particularly advantageous when  surrogate expansions are expected to be sparse, i.e. the QoI can be accurately represented with a few polynomial chaos (PC) basis functions.  Compressive sampling was first introduced in the field of signal processing to recover sparse signals using a number of samples  significantly smaller that the conventionally used Shannon-Nyquist sampling rate \cite{candes2008introduction,donoho2006compressed,candes2006robust}.  Motivated by the fact that the solution of many high dimensional problems of interest, such as high dimensional PDEs, can be represented by sparse, or at least approximately sparse, PCEs, CS was proposed in \cite{doostan2011non,mathelin2012compressed,blatman2011adaptive} to estimate PC coefficients in underdetermined cases. As  CS theorems suggest, the success of sparse estimation of PCE depends not only upon the sparsity of the solution of stochastic system, but  also on the coherence property of the Vandermonde-like  measurement matrix, formed by evaluations of orthogonal polynomials at sample locations \cite{candes2008introduction}, as will be elaborated later.
 
 Several efforts have been made in order to improve the two mentioned conditions for successful recovery. For instance in \cite{yang2016enhancing}, for Hermite expansions with Gaussian input variables, the original inputs are rotated such that a few of the new coordinates, i.e. linear combinations of original inputs, have significant impact on QoI, thereby increasing the sparsity of solution and, in turn,  the accuracy of recovery. The  second condition, i.e. coherence of the measurement matrix,  can be poor especially when trial expansions are high-order and/or high-dimensional. To remedy this, the  iterative approaches in \cite{jakeman2015enhancing, alemazkoor2016divide} can be used to optimally include only the ``important" basis functions into the trial expansion and its associated measurement matrix. Focusing on this second condition, another class of methods have  proposed sampling strategies that produce  less coherent measurement matrices \cite{rauhut2012sparse,tang2014subsampled,hampton2015compressive}. Among these approaches, the sampling strategy proposed in \cite{hampton2015compressive} was designed to be optimal in achieving the lowest local-coherence. 
 
 In this work, we introduce a near-optimal sampling strategy by further improving the local-coherence-based sampling of \cite{hampton2015compressive} and filtering  sample locations based on  cross-correlation properties of the resulting measurement matrix. Specifically, we establish  quantitative measures to capture these  cross-correlation properties between measurement matrix  columns, and use these measures as the criteria for near-optimal identification of sample locations.  It will be demonstrated that a sampling strategy that seeks to optimize these measures will lead to CS results that on average outperforms all other CS sampling strategies. This paper is organized as follows. Section~\ref{sec:CS in UQ} presents general concepts in compressive sampling and its theoretical background. In Section~\ref{sec:optimal sampling}, we introduce our sampling algorithm along with relevant theoretical supports. Finally, Section~\ref{sec:numericalresults} includes numerical examples and discussions about the advantages of the proposed approach.

\section{Setup and background} \label{sec:CS in UQ}

\subsection{Polynomial chaos expansion}
Let $I_{\bm \Xi} \subseteq \mathbb{R}^d $ be a tensor-product domain that is the support of $\bm \Xi$, where $\bm \Xi=(\Xi_1, ..., \Xi_d)$ is the vector of independent random variables, i.e. $\Xi_i \in I_{ \Xi_i}$ and $I_{\bm \Xi} = \times_{i=1}^d I_{ \Xi_i} $. Also, let $\rho_i: I_{\Xi_i} \rightarrow \mathbb{R}^+$ be the probability measure for variable $\Xi_i$ and let $\rho(\bm \Xi)=\prod_{i=1}^{d}\rho_i(\Xi_i)$. Given this setting, the set of univariate orthonormal polynomials, $\{ \psi_{\alpha,i} \}_{ \alpha \in \mathbb{N}_0}$ , satisfies

\begin{equation} \label{eq:orthonormality1d}
\int_{I_{\Xi_i}} \psi_{ \alpha,i}(\xi_i) \psi_{ \beta,i}(\xi_i)  \rho_i({\xi_i}) \diff  \xi_i = \delta_{\alpha \beta}, \quad \alpha,\beta \in \mathbb{N}_0.
\end{equation}
where $\mathbb{N}_0 = \mathbb{N} \cup \{0\}$, and $\delta_{{mn}}$ is the delta function. Therefore, the density function of $\Xi_i$, $\rho_i(\Xi_i)$, determines the type of polynomial. For example, Gaussian and uniform probability distributions enforce Hermite and Legendre polynomials, respectively.  The $d$-dimensional orthonormal polynomials are then derived from the multiplication of one dimensional polynomials in all dimensions. For example,
\begin{equation}
 \psi_{\bm \alpha}(\bm \xi)= \psi_{\alpha_{1},1}(\xi_1)\psi_{\alpha_{2},2}(\xi_2) ... \psi_{\alpha_d,d}(\xi_d), \quad  \bm \alpha=(\alpha_{1},\alpha_{2},..., \alpha_{d}).       
\end{equation}
Consequently, we have
\begin{equation} \label{eq:orthonormality}
\int_{I_{\bm \Xi}} \psi_{\bm \alpha}(\bm \xi) \psi_{\bm \beta}(\bm \xi)  \rho({\bm \xi}) \diff \bm \xi = \delta_{\bm\alpha\bm\beta}, \quad {\bm\alpha,\bm\beta} \in \mathbb{N}_0^d.  
\end{equation}
Using this construction, any  function $u(\bm \Xi):I_{\bm \Xi} \rightarrow \mathbb{R}$ that is square-integrable can be represented as

\begin{equation}\label{eq:repUasSum}
u(\bm{\Xi})= \sum_{\bm \alpha \in \mathbb{N}_0^d} c_{\bm \alpha } \psi_{\bm \alpha}(\bm\Xi), 
\end{equation}
where   $\{ \psi_{\bm \alpha} \}_{\bm \alpha \in \mathbb{N}_0^d}$ is the set of orthonormal basis functions satisfying Equation (\ref{eq:orthonormality}). 
However, for computation's sake, $u(\bm \Xi)$ is approximated by a finite order truncation of PC expansion given by 
\begin{equation}\label{eq:PCEexpansion}
u_{k}(\bm\Xi) := \sum_{\bm \alpha \in \Lambda_{d,k}} c_{\bm \alpha } \psi_{\bm \alpha}(\bm \Xi),
\end{equation}
where $k$ is the total order of the polynomial expansion and $\Lambda_{d,k}$ is the set of multi-indices defined as
\begin{equation} \label{eq:basisSet}
\Lambda_{d,k} := \{\bm \alpha \in \mathbb{N}_0^d : \Vert \bm \alpha \Vert_1 \le k \}.
\end{equation}
The cardinality of $\Lambda_{d,k}$, i.e. the  number of expansion terms, here denoted by $K$, is a function of $d$ and $k$ according to
\begin{equation}\label{eq:NumOfbasis}
K:= \vert \Lambda_{d,k} \vert = \frac{(k+d)!}{k!d!}.
\end{equation}
Given this setting, $u_{k}(\bm \Xi)$ approximates $u(\bm \Xi)$ in a proper sense and is referred to as the $k$-th degree PC approximation of $u(\bm \Xi)$. Each of the $K$ coefficients involved in the definition of $u_{k}$ can be exactly calculated by projecting $u(\bm \Xi)$ onto the associated basis function:

\begin{equation}\label{eq:exactCoeff}
c_{\bm \alpha^j} = \frac{1}{\gamma_{\bm \alpha^j}}\int_{I_{\bm \Xi}}u(\bm \xi)\psi_{\bm \alpha^j}(\bm \xi) \rho(\bm \xi) \diff \bm \xi,  \quad \bm \alpha^j \in \Lambda_{d,k}  =  \{\bm \alpha^1, \cdots, \bm \alpha^K \}. 
\end{equation}

Numerical approaches such as quadrature rule or sparse grid are usually used to approximate the above integral as its exact evaluation might be very cumbersome or not even possible \cite{eldred2008evaluation,xiu2010numerical}. However, as the dimensionality of the problems increases, the number of samples required by these numerical techniques for an accurate integral evaluation  increases exponentially. Supplying such a large sample size is prohibitively costly, especially if these samples are drawn from expensive high-fidelity simulations or costly experiments. As a result, efforts have been made to develop adaptive approaches that reduce the number of required samples by placing fewer samples on dimensions that do not impact QoI significantly \cite{hegland2003adaptive, ma2009adaptive}.  Another widely used approach to estimate PC coefficients is linear regression. In order to use linear regression, the number of samples must be equal or greater than the number of PC expansion terms. The well accepted oversampling rate is around 1.5 to 3 times the number of unknown coefficients. Recently, a quasi-optimal sampling approach was introduced in  \cite{shin2016nonadaptive} that results in accurate regression with $\mathcal{O}(1)$ oversampling. 

As briefly mentioned in Section \ref{sec:intro}, when the QoI is sparse with respect to the PC basis functions, compressive sampling approach have proved effective in estimating expansion coefficients using a number of samples that is significantly smaller than the number of  PC terms.  This section follows with the review on the basics of sparse PC recovery using compressive sampling. 

\subsection{Stochastic collocation using compressive sampling}
Compressive sampling was first introduced in the field of signal processing, where conventionally the number of samples required to recover a signal was determined by Shannon-Nyquist sampling rate \cite{candes2008introduction}. Compressive sampling allows for successful signal recovery using significantly fewer samples when the signal of interest is sparse. Therefore, it has been extensively applied in cases where the number of available samples is limited \cite{lustig2005application, ender2010compressive, paredes2007ultra, gemmeke2010compressive}. Due to this very advantage, compressive sampling has recently gained substantial attention in UQ, more specifically in stochastic collocation, where a sample set is used to build an analytical surrogate model (typically in the form of a PC expansion) \cite{doostan2011non, mathelin2012compressed, hampton2015compressive, yan2012stochastic, yang2013reweighted}. In what follows, a formal brief  background on the use of compressive sampling for PCE estimation is provided.

The objective of PCE estimation is to calculate the vector of unknown expansion coefficients $\bm c= (c_{\bm \alpha^1}, .., c_{\bm \alpha^K})^T$ in Equation~(\ref{eq:PCEexpansion}), given $M$ samples of the QoI, denoted by the data vector $\bm u=(u (\bm \xi^{(1)}),...,u (\bm \xi^{(M)}))^T$. These sampled outputs are evaluated at the $M$ realizations, $\left \{ \bm \xi^{(i)} \right \}_{i=1}^M$,  of model input $\bm \Xi$. Requiring $u_k(\bm \Xi)$ to approximate $u(\bm \Xi)$ results in the following system of equation,
\begin{equation}
\bm \Psi \bm c= \bm u,
\end{equation}
where  $\bm \Psi$ is the measurement matrix, constructed according to 
\begin{equation} \label{mesmatrix}
\bm \Psi=[\psi_{ij}], \quad \psi_{ij}=\psi_{\bm \alpha^{j}}(\bm \xi ^{(i)}), \quad 1\leqslant i\leqslant M, \quad  1\leqslant j\leqslant K. 
\end{equation}

We are interested in the underdetermined case where $M\leqslant K$. Under such condition, there may exist infinity many solutions for $\bm c$. Compressive sampling approach can be readily borrowed to find the sparsest solution, by formulating the sparse recovery problem as
\begin{equation} \label{eq:l0min}
\underset{\bm c}{\textrm{min}} \left \| \bm c \right \|_{0} \quad \textrm{subject to} \quad  \bm \Psi \bm c= \bm u,
\end{equation}
where $\left \| \cdot \right \|_0 $ indicates the $\ell_0$-norm, i.e. the number of non-zero terms. However, since $\ell_0$-norm is non-convex and discontinuous, the above problem is NP-hard. Therefore, $\ell_0$ minimization is usually replaced by its convex relaxation where $\ell_1$-norm of the solution $\bm c$ is minimized instead, i.e.
\begin{equation} \label{eq:l1min}
\underset{\bm c}{\textrm{min}} \left \| \bm c \right \|_{1} \quad \textrm{subject to} \quad  \bm \Psi \bm c= \bm u,
\end{equation}
$\ell_1$ minimization is the closest convex problem to $\ell_0$ minimization. It has been shown that when $\bm \Psi$ is sufficiently incoherent, as will be explained later, and $\bm c$ is sufficiently sparse, the solution of $\ell_0$ minimization is unique and is equal to the solution of $\ell_1$ minimization \cite{bruckstein2009sparse}. The minimization in (\ref{eq:l1min}) is named basis pursuit \cite{chen2001atomic}, and it can be solved using linear programming. If the measurements are known to be noisy, the problem can be reformulated as
\begin{equation} \label{eq:l1minerror}
\underset{\bm c}{\textrm{min}} \left \| \bm c \right \|_{1} \quad \textrm{subject to} \quad  \left \|\bm \Psi \bm c - \bm u  \right \|_2 < \epsilon,
\end{equation}
which is known as the basis pursuit denoising problem. $\epsilon$ controls the accuracy tolerance and can be prescribed if the distribution of measurement noise  is known. For example, if the measurement noise follows a normal distribution with zero mean and standard deviation $\sigma_n$, then $\epsilon$ is naturally set to be $\sqrt{M\sigma_n^2}$. When instead of measurement samples, simulation samples are used in the PC estimation, and a distribution for numerical or modeling error is absent, one may use cross-validation  to determine the tolerance parameter  $\epsilon$. This section continues with a brief review on  theorems developed on recoverability of compressive sampling and efforts heretofore made to improve the accuracy of   PC estimation using compressive sampling.  

\subsection{PC recoverability using compressive sampling} \label{gPC recovery}
Whether desired solution can be  recovered from compressive sampling or not mainly depends on the properties of the measurement matrix $\bm \Psi$. One of the properties that are shown to be significantly relevant is the Restricted Isometry Property (RIP) or the $s$-restricted isometry constant for the measurement matrix \cite{candes2008restricted}. Formally speaking  the $s$-restricted isometry constant for a matrix $\bm \Psi \in \mathbb{R}^{M \times K}$ is defined to be the smallest $\delta_{s} \in (0,1)$ such that
\begin{equation} \label{eq:RIP}
(1-\delta_{s})\left \| \bm c \right \|_{2}^{2}\leqslant  \left \| \bm \Psi^{s^{*}}\bm c\right \|_{2}^{2}\leqslant  (1+\delta_{s})\left \| \bm  c \right \|_{2}^{2},
\end{equation}
for every submatrix  $\bm \Psi^{s^{*}} \in \mathbb{R}^{M \times s^{*}},  s^{*} \leqslant s$ , of $\bm \Psi$ and every vector $\bm c \in \mathbb{R}^{s^{*}}$. Thus defined, a small RIP constant for a measurement matrix will lead to a situation wherein for any given sparse signal, energy of the measured signal is not very different from the energy of the given signal itself; hence the appeal of a small RIP constant. The following theorem provides an upperbound on RIP constant so that the optimization problem (\ref{eq:l1minerror})  leads to accurate recovery.

\begin{theorem}[\cite{candes2008restricted}]\label{thm.rip}
Let $\bm \Psi \in \mathbb{R}^{M\times K}$ with RIP constant $\delta_{2s}$ such that $\delta_{2s}<\sqrt{2}-1$. For a given $ \bar{\bm c}$, and noisy measurement $\bm y=\bm\Psi \bar{\bm c}+ \bm \epsilon$, $\left \| \bm \epsilon \right \|_2\leqslant \epsilon$,  let $\bm c$ be the solution of 
\begin{equation} \label{eq:l1minerror1}
{\textrm{min}}\left \| \bm c \right \|_{1} \quad \textrm{subject to} \quad  \left \|\bm \Psi \bm c - \bm y \right \|_2 < \epsilon .
\end{equation}
Then the reconstruction error satisfies 
\begin{equation}
\left \| \bm c-\bar{\bm c} \right \|\leqslant C_1\frac{\left \| \bm c- \bm c^{*} \right \|}{\sqrt{s}} + C_2\epsilon,
\end{equation}
where  $C_1$ and $C_2$ only depend on $\delta_{2s}$ and $\bm c^*$ is the vector $\bm c$ with all but the s-largest entries set to be zero. If $\bm c$ is s-sparse and the measurements are noiseless, then the recovery is exact.  
\end{theorem}

Calculating RIP constant for a given matrix is an NP-complete problem \cite{maleki2010approximate}. The following  theorem  gives a probabilistic upperbound on RIP constant for bounded orthonormal systems. Let $\{ \psi_{n}\}_{1\leqslant n \leqslant K} $ be an orthonormal bounded system of functions on $\mathcal{D}$, where $\mathcal{D}$ is endowed with a probability measure $\nu$. Specifically, we have
\begin{equation} \label{eq:orthogonalitygeneral}
\int_{\mathcal{D}} \psi_{ n}( \xi) \psi_{m}(\xi)  \nu({\xi}) \diff \xi  =\delta_{mn}, \quad  {1\leqslant m, n \leqslant K},  
\end{equation}
and the following uniform bound,
\begin{equation} \label{boundedorthogonalgeneral}
\left \| \psi_{n} \right \|_{\infty}= \ \underset{  \xi \in \mathcal{D}}{\textrm{sup}}
\left | \psi_{n}( \xi) \right | \leqslant L, \quad \text {for all} \quad {1\leqslant n \leqslant K}.
\end{equation}
Consequently, an orthonormal polynomial system $\{ \psi_{\bm \alpha} \}_{\bm \alpha \in \Lambda _{d,k}}$ defined in (\ref{eq:orthonormality}) is a bounded orthonormal system if it is uniformly bounded by 
\begin{equation} \label{boundedorthogonal}
\underset{ \bm \alpha \in \Lambda _{d,k}}{\textrm{sup}}\left \| \psi_{\bm \alpha} \right \|_{\infty}=\underset{ \bm \alpha \in \Lambda _{d,k}}{\textrm{sup}} \ \underset{ \bm \xi \in \Omega}{\textrm{sup}}
\left | \psi_{\bm \alpha}(\bm \xi) \right | \leqslant L,
\end{equation}
for some constant $L\geq1$.  Hereinafter, let us refer to the bound $L$ as the local-coherence.

\begin{theorem}[\cite{rauhut2010compressive}]\label{thm.coherencePCE}
	Let $\{\psi_{n}\}_{1\leqslant n \leqslant K}$ be a bounded orthonormal system satisfying Equations~(\ref{eq:orthogonalitygeneral}) and  (\ref{boundedorthogonalgeneral}). Also let  $\bm \Psi \in \mathbb{R}^{M \times K}$ be a measurement matrix with entries $\left \{\psi_{ij}=\psi_{j}( \xi ^{(i)})\right \}_{1\leqslant i\leqslant M, 1\leqslant j\leqslant K}$, where $\xi ^{(1)}, ..., \xi ^{(M)}$ are random samples drawn from measure $\nu$. Assuming that 
	\begin{equation}
	M\geqslant C\delta^{-2}L^2s\log^3(s)\log(K),
	\end{equation}
then with probability at least $1-K^{\beta \log^3(s)}$ the RIP constant $\delta_s$ of $\frac{1}{\sqrt{M}}\bm\Psi$ satisfies $\delta_s \leqslant \delta$. The $C, \beta > 0 $ are universal constants. 
\end{theorem}
 
 It is worthwhile to highlight the differences between these two theorems. Theorem~\ref{thm.rip} is a deterministic theorem, while Theorem~\ref{thm.coherencePCE} is probabilistic. Theorem~\ref{thm.rip} provides a deterministic guarantee that is universal, in the sense that once a specific RIP property is satisfied for samples of a class of measurement matrices, then the corresponding error bounds hold for all kinds of target expansions and recovery is always exact for cases with noiseless measurement and $s$-sparse target. On the other hand, when the condition of Theorem~\ref{thm.coherencePCE} is satisfied,  i.e. sufficient number of samples are provided, the recovery accuracy can be achieved with high probability, and not  with probability 1.

\section{Near-optimal sampling strategy} \label{sec:optimal sampling}
Suppose that a limited sampling budget allows for $M$ samples to be drawn. Our objective, then, is to optimally select $M$ locations in the parameter space, at which samples of QoI are  computationally or experimentally evaluated. Let us pose a slightly different, but closely related, question: out of a sufficiently large pool of $M_p$ candidate sample locations, how can one  identify the $M$ ``optimal"  locations? The sufficiently large number $M_p$ can be determined such that an accurate regression for the given stochastic system is achieved. In what follows,  we first discuss the sampling strategies that aim at improving the local-coherence. This will then be followed by discussions about  other relevant properties, in particular those of the measurement matrix, that have implications on accuracy and stability of sparse recovery, and how a `near-optimal' sampling strategy can be designed to improve these relevant properties.  
 
 \subsection{Sampling strategies focusing on  local-coherence}\label{sec.samplingstrategies}
As is evident by the theoretical results of Section~\ref{gPC recovery}, specifically in Theorem \ref{thm.coherencePCE},  local-coherence  has an implication on recovery accuracy. This has motivated researchers to   develop new sampling strategies such that this property is improved. The conventional sampling approach, namely the standard (random) sampling approach,  involves drawing random realizations from uniform and normal distributions for Legendre and Hermite PC expansions, respectively. In \cite{rauhut2012sparse}, it was recommended to use a preconditioned Legendre measurement matrix, where samples are taken from Chebyshev distribution, instead of a uniform distribution, to improve accuracy.  With respect to this new sampling distribution, new orthogonal basis functions were calculated by scaling  or weighting the Legendre polynomials. It was shown in   \cite{yan2012stochastic} that when system dimensionality $d$ is larger than  polynomial order $k$, the Chebyshev preconditioning approach will result in a polynomial system with a larger local-coherence, $L$, and subsequently in less accurate results compared with that from standard  sampling. An alternative sampling strategy was proposed in \cite{tang2014subsampled}, where the Legendre polynomial system was first turned into a discrete orthogonal system using Gauss quadrature points, from which  samples are drawn randomly for sparse recovery. But, it was shown that the new discretized orthogonal system improves the bound $L$ only when $k\geqslant d$, and for $k < d$,  recovery accuracy will be lowered than that from standard sampling. 

In \cite{hampton2015compressive}, a coherence-optimal, or to be compliant with the terminology in this paper, a local-coherence-optimal sampling approach was introduced, where instead of directly sampling from the probability measure, $\rho(\bm \xi)$, samples are drawn from a different ``optimal" probability measure, $\rho_{o}(\bm \xi)$, constructed according to 
 \begin{equation}
 \rho_{o}(\bm \xi)= C^2 \rho(\bm \xi) B^2(\bm \xi),
 \end{equation}
 where $C$ is a normalizing constant and 
 \begin{equation}
 B(\bm \xi):= \underset{\bm \alpha\in \Lambda _{d,k}}{\textrm {max}}\left |  \psi_{\bm \alpha}(\bm \xi)\right |. 
 \end{equation}
 Corresponding to this probability measure, the weight function, used to retrieve the orthogonality of the basis functions, should be 
 \begin{equation} \label{weight}
 w(\bm \xi)=\frac{1}{B(\bm \xi)}. 
 \end{equation}
 Accordingly, the $\ell_1$-minimizations in (\ref{eq:l1min}) and (\ref{eq:l1minerror}) were adjusted to include the weight functions. The following weighted $\ell_1$-minimizations were then used for sparse recovery
 \begin{equation} \label{eq:weightedl1min}
 \underset{\bm c}{\textrm{min}} \left \| \bm c \right \|_{1} \quad \textrm{subject to} \quad  \bm W \bm \Psi \bm c= \bm W \bm u,
 \end{equation}

 \begin{equation} \label{eq:weightedl1minerror}
 \underset{\bm c}{\textrm{min}} \left \| \bm c \right \|_{1} \quad \textrm{subject to} \quad  \left \|\bm W \bm \Psi \bm c - \bm W \bm u  \right \|_2 < \epsilon,
 \end{equation}
where $\bm W$ is the $M \times M$ diagonal weight matrix, with $\bm W(i,i)= w(\bm \xi^{(i)})$ for $i=1,\cdots, M$.  It was also proved in \cite{hampton2015compressive} that this sampling approach results in the lowest local-coherence among all sampling approaches, and can  thus outperform standard sampling for both cases, i.e., $k<d$ and $k\geqslant d$. 

In what follows, we discuss how this strategy can be further improved by considering other relevant properties with implications on recovery accuracy.

\subsection{Sampling strategies focusing on measurement matrix properties}\label{optimal sampling in cs}
It is obvious that to recover all $s$-sparse vectors $\bar{\bm c} $ from observation vector $\bm u=\bm \Psi \bar{\bm c}$, using the $\ell_0$ minimization problem in (\ref{eq:l0min}), we must have $\bm \Psi \bar{\bm c} \neq \bm \Psi  \bm c' $ for any pair of distinct $s$-sparse vectors $\bm c', \bar{\bm c}$. In other words, we want $\bm c' \neq \bar{\bm c}$ to lead to distinct observation vectors. For this to happen, the measurement matrix should not satisfy $\bm \Psi ( \bar{\bm c} - \bm c') = 0 $, for any vector  $\bar{\bm c} -   \bm c' $ that  has 2$s$ or fewer nonzero terms. This means that the null space of $\bm \Psi$ should  contain no vector with $2s$ or fewer nonzero terms. In compressive sampling, this property is mostly characterized using \textit{spark} \cite{davenport2011introduction}. The spark of a matrix is the smallest number of its columns that are linearly dependent. Using the definition of spark, it is then straightforward to guarantee that the solution of $\ell_0$ minimization given in (\ref{eq:l0min}) exactly recovers the $s$-sparse $\bar{\bm c} $ if   \text{spark}$(\bm \Psi)> 2s$ \cite{donoho2003optimally}. Moreover, it can be shown that in this case the $\ell_1$ minimization  in (\ref{eq:l1min}) also recovers the exact solution, $\bar{\bm c} $ \cite{li2013projection}. 

Clearly, a measurement matrix with a larger spark allows exact recovery using compressive sampling for a larger set of target signals (solution vectors). Therefore, it is desired to maximize the spark of measurement matrices. However, computing the spark of a matrix is an NP-hard problem \cite{cheng2015fundamentals}. As an alternative, one can analyze recovery guarantees using alternative properties which are easier to compute. One such property is the mutual-coherence, which for a given matrix $\bm \Psi \in \mathbb{R}^{M \times K}$ is defined as the maximum absolute normalized inner product, i.e. cross-correlation, between its columns \cite{doostan2011non, donoho2006stable}.  Let $\psi_{1}, \psi_{2}, ..., \psi_{K} \in \mathbb{R}^{M}$ be the columns of matrix $\bm \Psi$. The mutual-coherence of matrix $\bm \Psi$, denoted by $\mu(\bm \Psi)$, is then given by
\begin{equation} \label{eq:coherence}
\mu(\bm \Psi):= \underset{1\leqslant  i, j\leqslant  K, i\neq j}{\max}\frac{|\psi_{j}^{T}\psi_{i}|}{\begin{Vmatrix}\psi_{j}\end{Vmatrix}_{2}\begin{Vmatrix}\psi_{i}\end{Vmatrix}_{2}}.
\end{equation}\\

 The following simple proposition explains how spark and mutual coherence are related. 
 \newtheorem{prop}{Proposition}
 \begin{prop}[\cite{elad2010sparse}] \label{prop1}
 For any matrix $\bm \Psi \in \mathbb{R}^{M \times K}$ the following holds:
 \begin{equation} \label{spark}
 \text{spark}(\bm \Psi)\geqslant 1+\frac{1}{\mu(\bm \Psi)}.
 \end{equation}
 \end{prop}
 
 Mutual-coherence is an indicator of the worst interdependence, i.e. maximum cross-correlation between the columns of a matrix. It is zero for an orthogonal matrix and is strictly positive when $M<K$.   Proposition \ref{prop1} makes it obvious that a small value for mutual-coherence  is desired. It may be concluded that measurement matrix $\bm \Psi$ should be designed in a way that its mutual-coherence, i.e. its maximum cross-correlation,  is minimized. However, it has been observed that minimizing maximum cross-correlation  does not necessarily improve the recovery accuracy of compressive sampling \cite{li2013projection,elad2007optimized}. This is because Equation (\ref{spark}) only provides a lower bound on the spark. Therefore, minimizing mutual-coherence considers only the worst-case scenario and fails to account for other possibilities for improving compressive sampling performance  \cite{elad2007optimized}. 
 
 Our objective in this work is to design a measurement matrix  $\bm \Psi$ which leads to better  compressive sampling performance \emph{on average}, and not merely in the worst-case scenario. To this end, we need to determine which property of the measurement matrix should be optimized. This property should ideally be one that directly and sufficiently controls the accuracy of compressive sampling. A widely-accepted property is the RIP constant, as suggested by Theorem \ref{thm.rip}. As previously mentioned, calculating RIP constant for a given matrix is an NP-complete problem. However, it is known that the eigenvalues of each column-submatrix  $\bm \Psi^{s^{*}} \in \mathbb{R}^{M \times s^{*}},  s^{*} \leqslant s$ can be bounded by the RIP constant, $\delta_s $ 
 \begin{equation}
 1-\delta_s \leqslant \lambda _{\min}(\bm \Psi^{s^{*}T}\bm \Psi^{s^{*}})\leqslant \lambda _{\max}(\bm \Psi^{s^{*}T}\bm \Psi^{s^{*}})\leqslant 1+\delta_s.
 \end{equation}
Therefore, one may suggest to minimize the condition number of all column-submatrices with $s$ or fewer columns. The challenge, however, is that the sparsity level of solution, $s$, is not known in advance. Moreover, calculating such a combinatorial measure is not a trivial task and can be computationally impossible \cite{elad2007optimized}. 
 
As a result, establishing a single matrix property that  sufficiently guarantees   the accuracy of compressive sampling method and is easily computable still remains an open challenge  \cite{li2013projection,elad2007optimized}. To sidestep this challenge, efforts have focused on identifying properties or measures that are relatively better than maximum cross-correlation, or mutual-coherence. In \cite{elad2007optimized}, for the first time, it was suggested that a $t$-averaged  mutual-coherence  be minimized. Denoted by $\mu_t(\bm \Psi)$, the $t$-averaged  mutual-coherence  for a measurement matrix $\bm \Psi$ is defined as the average of cross-correlations larger than  threshold $t$, i.e.
 \begin{equation} \label{eq:avecoherence}
 \mu_t(\bm \Psi)=\frac{\underset{1\leqslant  i, j\leqslant  K, i\neq j}{\sum} \mathbbm{1}(\left | g_{ij} \right |\geqslant t).\left |  g_{ij}\right | }{\underset{1\leqslant  i, j\leqslant  K, i\neq j}{\sum} \mathbbm{1}(\left | g_{ij} \right |\geqslant t)},
 \end{equation}
where $\mathbbm{1}(\cdot)$ is the indicator function, $g_{ij}$ is the $ij$th component of the Gram matrix  $\bm G= \tilde{\bm \Psi}^T\tilde{\bm \Psi}$,  and $\tilde{\bm \Psi}$ is the column-normalized version of $\bm \Psi$. It has been shown that recovery accuracy can be significantly improved if instead of a random measurement matrix, the measurement matrix is optimized based on $\mu_t(\bm \Psi)$. However, it was later shown in \cite{duarte2008learning} that  a $\mu_t(\bm \Psi)$-optimized measurement matrix is not robust when  target signals are attached to some noise, and as such,  not exactly sparse. 

In order to improve this robustness, efforts have focused on optimizing measurement matrices by considering all the cross-correlations, and not the ones larger than a  threshold. Specifically, these efforts seek to minimize the distance between the Gram matrix, $\bm G$, and the corresponding identity matrix. With $\left \|  \cdot \right \|_F$ denoting the Frobenius norm, it has been shown that the measurement matrix determined by solving the following minimization problem
\begin{equation}\label{eq:orthomin}
\min_{\bm \Psi \in R^{M \times K}} \left \|I_K - \tilde{\bm \Psi}^T\tilde{\bm \Psi} \right \|_{F}^2,
\end{equation}
can lead to a signal recovery that is both more accurate and more robust compared with that using the $\mu_t(\bm \Psi)$-optimized  matrix \cite{li2013projection,abolghasemi2010optimization, zelnik2011sensing, tian2016orthogonal}. 
In what follows, we propose our near-optimal sampling strategy which considers measures of maximum and average cross-correlation of measurement matrices, and optimizes sample locations accordingly. 

\subsection{Near-optimal sampling strategy}

To achieve a better robustness, as argued earlier, we aim to select  sample locations that collectively minimize a measure of average cross-correlation given by
\begin{equation} \label{eq:orthomes}
\gamma(\bm \Psi):= \frac{1}{N}\left \|I_K - \tilde{\bm \Psi}^T\tilde{\bm \Psi} \right \|_{F}^2,
\end{equation}
where $N:=K \times (K-1)$ is the total number of column pairs. For the sake of brevity, we refer to $\gamma(\bm \Psi)$ as the \emph{average cross-correlation}, even though precisely speaking, it is the average of squares of cross-correlation. It should be noted that minimizing this average cross-correlation alone will not necessarily result in the smallest maximum cross-correlation, and the  mutual-coherence could still be undesirably large and the sparse recovery  significantly inaccurate. As a remedy, we seek to minimize both the average cross-correlation $\gamma(\bm \Psi)$ and the mutual-coherence $\mu(\bm \Psi)$, simultaneously in a multi-objective problem. 

To solve the optimization problem in (\ref{eq:orthomin}) for PC measurement matrix, an option is to adopt the solution approaches proposed for (\ref{eq:orthomin}). These are mainly iterative approaches based on  gradient descent method \cite{abolghasemi2010optimization} or  bound-optimization method  \cite{zelnik2011sensing}. However, these algorithms are applicable  for measurement matrices that are less restricted compared to PC measurement matrix, in that their components are not constrained to be  evaluations of specific orthogonal basis functions at certain sample locations, and as such can be freely optimized. Therefore, a greedy algorithm is the natural option for solving (\ref{eq:orthomin}). Our greedy algorithm for  near-optimal sampling strategy begins by populating a large pool of candidate sample locations in the multidimensional parameter space, and seeks to find the near-optimal $M$ locations and the corresponding $M \times K$ measurement matrix. We select the first sample location randomly from the candidates pool, and this constitutes the first row in the measurement matrix.  At each step of the algorithm, a new sample location, or matrix row is added. This is done by searching through the candidates pool  for the best sample location.  In this two-objective  optimization problem, we select the ``best" sample location as the one with the smallest normalized distance with respect to the utopia point \cite{zavala2012stability,roman2006evenly}. The utopia point is an abstract point in our two-dimensional objective space, whose first and second coordinates  are the smallest maximum  cross-correlation and smallest average cross-correlation. Note that this abstract point is typically not among the Pareto optimal solutions of the two-objective optimization problem, and is therefore not attainable as an optimal solution.  This is why we select from the sample locations pool the  location that is closest to the utopia point, i.e. the sample location whose corresponding average cross-correlation and maximum cross-correlation have the smallest normalized distance with the coordinates of the utopia point, as is elaborated in the following pseudo-code.

Let $\bm \Psi^{\text{pool}}$ denote the $M_p \times K$ measurement matrix associated with the large pool of $M_p$ candidate locations, and  $\bm \Psi^{\text{opt}[M]}$ the near-optimal $M \times K$ row-submatrix of the ``pool" matrix $\bm \Psi^{\text{pool}}$. This submatrix is identified   through incremental row-concatenation, as shown in  Algorithm \ref{alg.greedy}. In this pseudo-code, $\bm \Psi^{\text{opt}[i]}$  of size $i \times K$ denotes the near-optimal submatrix at the $i$th step of the algorithm, and $\bm \Psi^{\text{pool}}_{(j)}$ represents the $j$th row in the ``pool" matrix.   Also, at each iteration, $\mu'_j$ and $ \gamma'_j$ are the maximum cross-correlation and average cross-correlation, respectively  recorded for  candidate location $j$.
 
 \begin{algorithm}[H] 
 	\caption{Greedy algorithm for near-optimal sample locations}\label{pseudocode}
 	\begin{algorithmic}[1]
 		\State Initiate $\bm \Psi^{\text{opt}[1]}$ to be a random row in $\bm \Psi^{\text{pool}}$
 		\For {$i=2:M$} 

 		\For {$j=1:M_p$}
 		\State  $\bm \Psi^{\text{temp}}$ = row-concatenate($\bm \Psi^{\text{opt}[i-1]},\bm \Psi^{\text{pool}}_{(j)})$ 
 	    \State $\mu'_j = \mu(\bm \Psi^{\text{temp}})$ and $\gamma'_j = \gamma(\bm \Psi^{\text{temp}})$
 		\EndFor
		\State $j^* = \underset{j}{\text{argmin}}\left [  \frac{\mu'_j-\text{min}(\bm \mu')}{\text{max}(\bm \mu')-\text{min}(\bm \mu')}\right ]^2+\left [ \frac{\gamma'_j-\text{min}(\bm \gamma')}{\text{max}(\bm \gamma')-\text{min}(\bm \gamma')} \right ]^2$ \Comment{$\bm \mu' = [ \mu'_1,\cdots,\mu'_{M_p}]$, $\bm \gamma'= [\gamma'_1,\cdots,\gamma'_{M_p}] $} 
			\State  $\bm \Psi^{\text{opt}[i]}$ = row-concatenate($\bm \Psi^{\text{opt}[i-1]},\bm \Psi^{\text{pool}}_{(j^*)})$ 
 		\EndFor
 		
 	\end{algorithmic}\label{alg.greedy}
 \end{algorithm}

As discussed in Section~ \ref{gPC recovery}, it has been shown that local-coherence-optimal sampling improves the PC coefficients recovery accuracy, over standard sampling. To exploit this advantage, we use the local-coherence-optimal sampling strategy to generate the large pool of candidate sample locations. In order to retain orthogonality,  we just need to substitute $\bm \Psi^{\text{pool}}$ with $\bm W \bm \Psi^{\text{pool}}$ in Algorithm \ref{pseudocode}, where $\bm W$ is the diagonal weight matrix defined earlier in Section~\ref{sec.samplingstrategies}. Compared to standard sampling and local-coherence-optimal sampling, our sampling strategy incurs extra computational cost due to additional row selections from the candidates pool in the greedy algorithm. However, this additional cost is typically negligible with respect to the cost of sample collection, especially when the benefit of improved accuracy is also considered.

We refer to the proposed sampling approach  as  near-optimal sampling strategy. The reason we do not use the term `optimal'  is two-fold: First,  
 even though studies in signal processing have shown significant improvement in recovery accuracy by minimizing (\ref{eq:orthomes}), this orthogonality property does  not by itself establish a sufficient  criterion for  recovery accuracy of compressive sampling (a measure that is both sufficient and tractable  has not been identified in the literature.) Second, our approach can be sensitive to (i) the random choice of candidate locations and (ii) the random choice of the initial location (first row in the submatrix), and is therefore not fully deterministic, and as such not optimal.  In the next section, we provide numerical examples to demonstrate advantages of our proposed sampling strategy.

\section{Numerical examples} \label{sec:numericalresults}
To demonstrate the advantage of the near-optimal sampling strategy over other sampling approaches, four target functions are considered in this section: (i) a low-dimensional high-order polynomial function, (ii) a high-dimensional low-order  polynomial function, (iii) a six-dimensional generalized Rosenbrock function, and (iv) the solution to a stochastic diffusion problem. This would allow for a comprehensive comparison of the sampling strategies for a variety of models with different combinations of dimension and order. The first three target functions are chosen to be exactly $s$-sparse (or sparse, in short), whereas the last one is  approximately $s$-sparse (or compressible, in short).  In all examples, optimization problems (\ref{eq:l1min}) and (\ref{eq:weightedl1min}), corresponding to a noise-less measurement setting, were formed and solved using the SPGL1 package \cite{van2007spgl1}.  Also, coherence-optimal samples were generated using  the `coh\_opt' package developed by the authors of \cite{hampton2015compressive}. For the sake of brevity, we only report the results for Legendre polynomial expansions in this paper, but note that similar improvements in recovery accuracy were observed when near-optimal sampling was applied to Hermite polynomial expansions. In all these examples, the proposed near-optimal sampling approach is compared with (i) standard sampling, and (ii) local-coherence-optimal sampling, or in short, `coherence-optimal' sampling.

\subsection{Low-dimensional high-order sparse PCE} \label{low-dim-high-order}
Let us consider the target function, $u(\bm \Xi)$, to be a sparse 20th-order Legendre polynomial expansion in a two-dimension random space with uniform density on $[-1,1]^{2}$, manufactured according to  
\begin{equation}\label{ex.lowDhighk}
u(\bm \Xi)= \sum_{i=1}^{5} \Xi_1^{2i}\Xi_2^{2i}.
\end{equation}
Using limited samples from this exact function, the objective is to recover the sparse coefficient vector. In selecting the sample locations to form the measurement matrix and data vector, the proposed near-optimal sampling approach is compared versus  random sampling strategies, namely  (i) standard sampling and (ii) coherence-optimal sampling. The candidates pool in the near-optimal sampling approach includes $100,000$ coherence-optimal samples. For all three approaches, we report the performance results obtained by 100 independent runs. This is to capture the variability induced by small sample size in standard and coherence-optimal sampling approaches. In the near-optimal approach, these independent runs  allow us to account for (i)  the sensitivity of performance with respect to the choice of initial sample location, and (ii) the variability induced by the finite size of the candidate pool, and (iii) the variability induced by the fact that our cross-correlation measures do not necessarily guarantee CS recovery accuracy.

\begin{figure} [H]
  	
  	\begin{subfigure}[t]{0.47 \linewidth}
  		\includegraphics[width=1\linewidth]{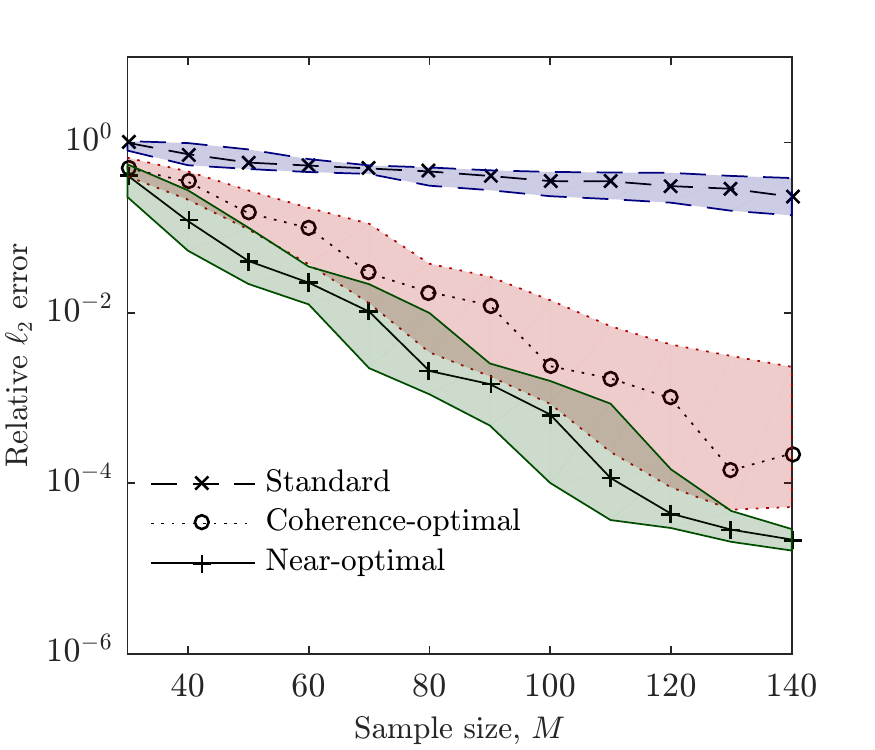}
  		\caption{}
  		\label{fig:example1-error-prcs}		
  	\end{subfigure}
  	\quad
  	\begin{subfigure}[t]{0.47 \linewidth}
  		\includegraphics[width=1\linewidth]{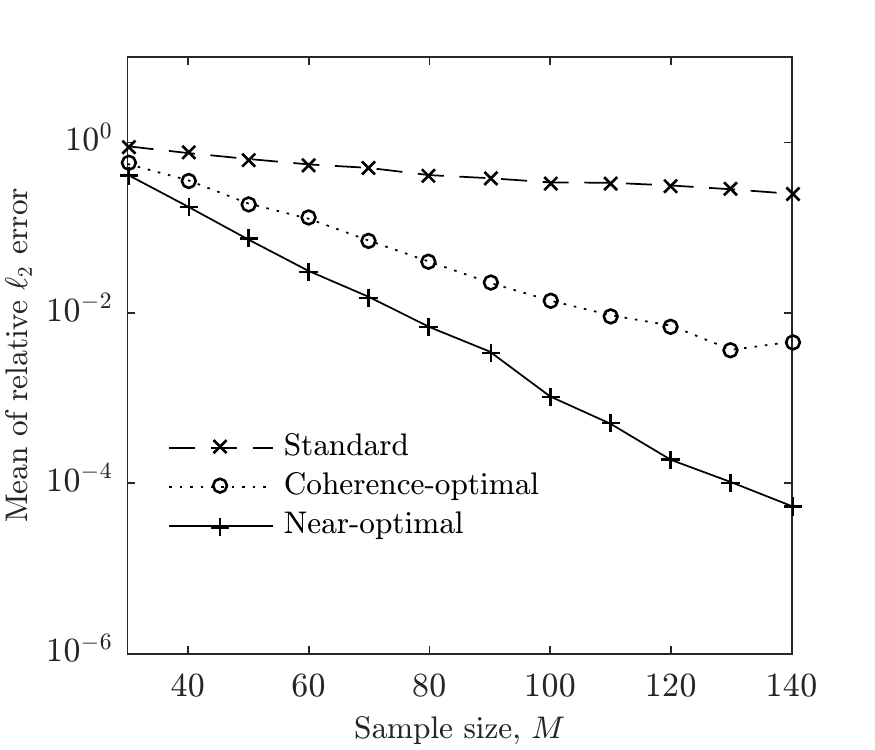}
  		\caption{}
  		\label{fig:example1-error-mean}
  	\end{subfigure}
  	
  	\begin{subfigure}[t]{0.47 \linewidth}
  		\includegraphics[width=1\linewidth]{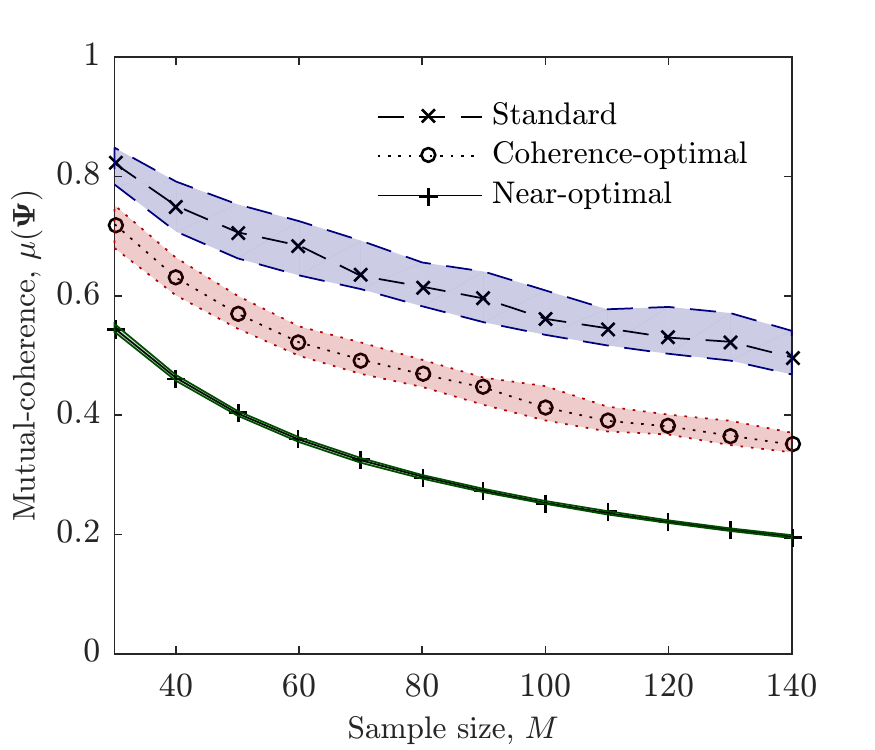}
  		\caption{}
  		\label{fig:example1-MuCoh}
  	\end{subfigure}
  	\quad
  	\begin{subfigure}[t]{0.47\linewidth}
  		\includegraphics[width=1\linewidth]{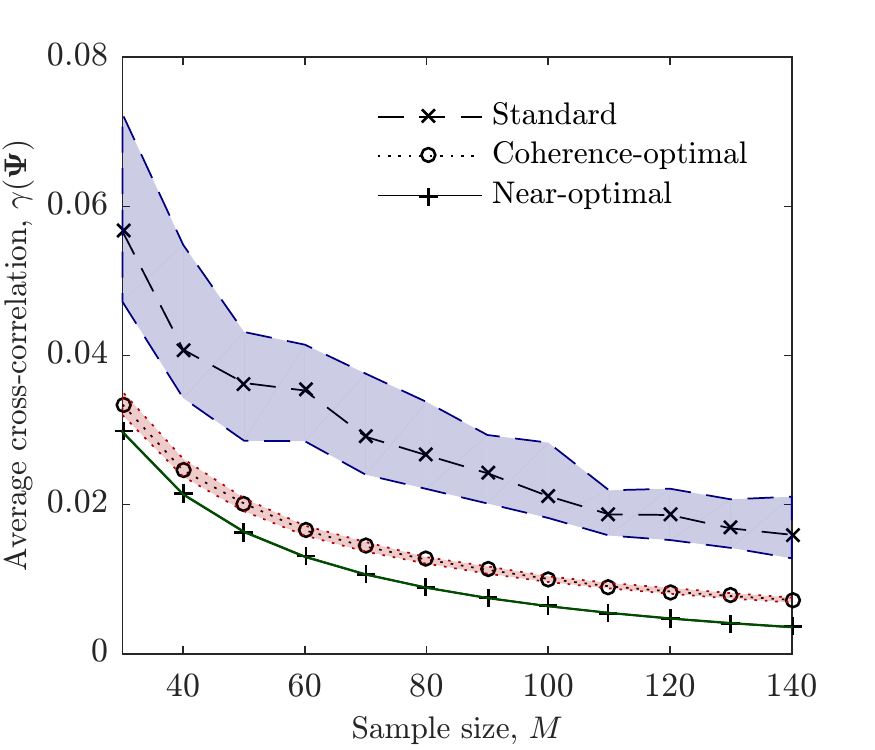}
  		\caption{}
  		\label{fig:example1-OrthMes}
  	\end{subfigure}

\caption{ Comparison of near-optimal sampling results with  standard and coherence-optimal  sampling methods for the low-dimensional high-degree manufactured PCE of Equation (\ref{ex.lowDhighk}). The following measures are compared: (a)  median, and 1st and 3rd quantiles of  relative $\ell_2$ error (b)  mean of relative $\ell_2$ error, (c) median,  and 1st and 3rd quantiles of mutual-coherence, and (d) median, 1st and 3rd quantiles of average cross-correlation.}
\label{fig.example1}
\end{figure}

Figure \ref{fig:example1-error-prcs} shows the median, and 1st and 3rd quantiles of relative $\ell_2$ error and Figure \ref{fig:example1-error-mean} shows the mean of relative $\ell_2$ error. The relative $\ell_2$ error is calculated as $\left \| \bm c - \bar{\bm c} \right \|_2 / \left \| \bar{\bm c} \right \|_2$, where $\bar{\bm c}$ is the exact coefficient vector and $\bm c$ is the solution of $\ell_1$ minimization in (\ref{eq:l1min}).  As is apparent in Figure \ref{fig:example1-error-prcs} and \ref{fig:example1-error-mean} ,  coherence optimal sampling results in a smaller error, compared to standard sampling, as the samples are drawn from a distribution with a smaller bound $L$ defined in (\ref{boundedorthogonal}). However, our proposed sampling outperforms the coherence-optimal one. To explain this, Figures \ref{fig:example1-MuCoh} and \ref{fig:example1-OrthMes} show the median, and 1st and 3rd quantiles of mutual-coherence and average cross-correlation for the measurement matrix, respectively. It can be seen that near-optimal sampling beats the other two approaches in both measures, leading to the higher observed accuracy in the PCE estimation.   

\subsection{High-dimensional low-order sparse PCE}

\begin{figure} 
  	
  	\begin{subfigure}[t]{0.47 \linewidth}
  		\includegraphics[width=1\linewidth]{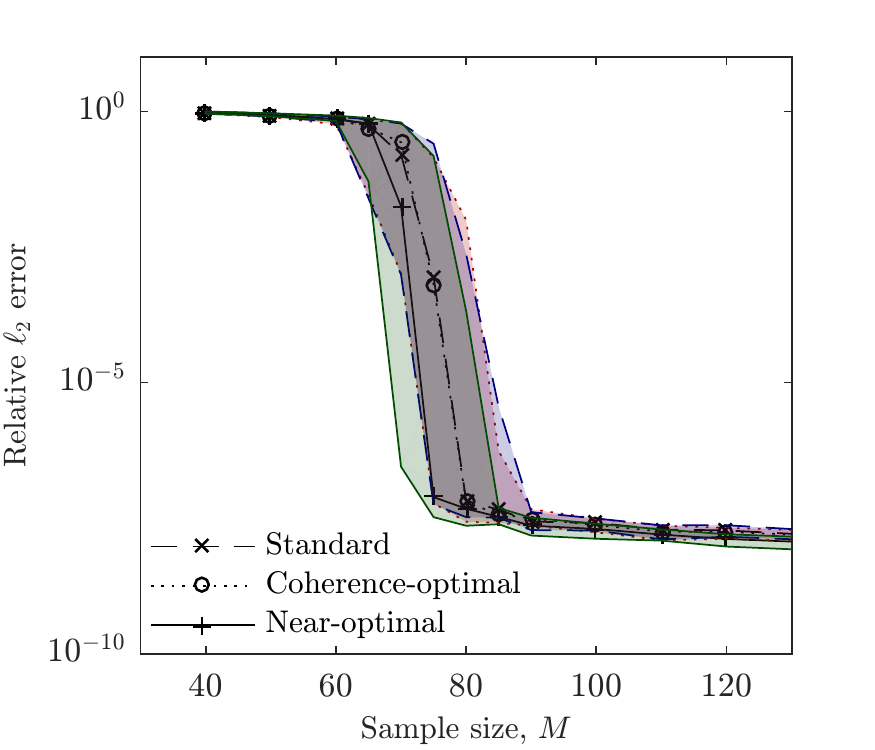}
  		\caption{}
  		\label{fig:example2-error-prcs}		
  	\end{subfigure}
  	\quad
  	\begin{subfigure}[t]{0.47 \linewidth}
  		\includegraphics[width=1\linewidth]{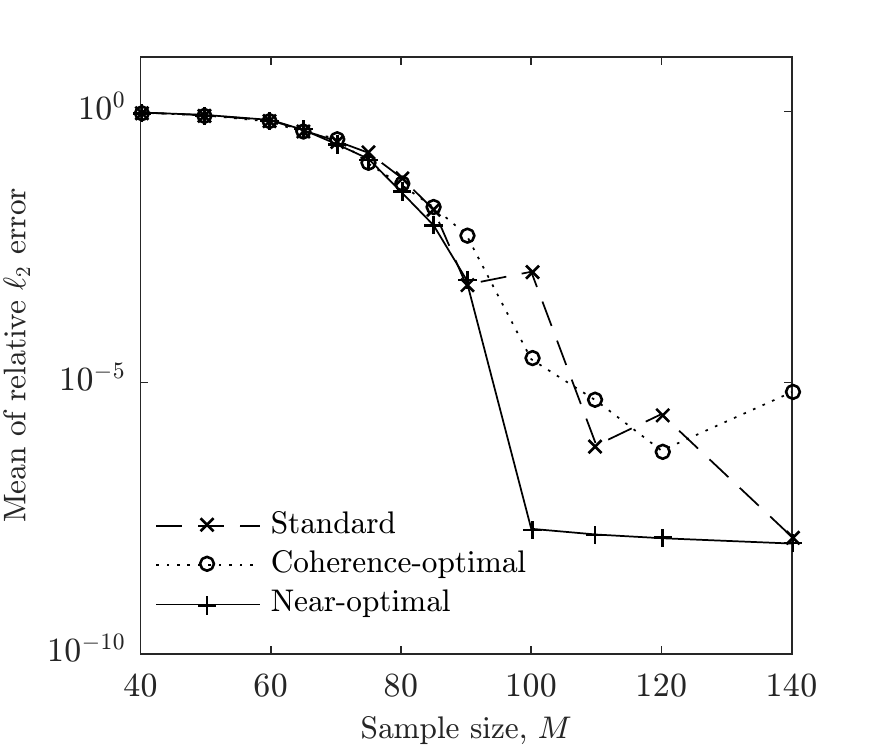}
  		\caption{}
  		\label{fig:example2-error-mean}
  	\end{subfigure}
  	 
  	\begin{subfigure}[t]{0.47 \linewidth}
  		\includegraphics[width=1\linewidth]{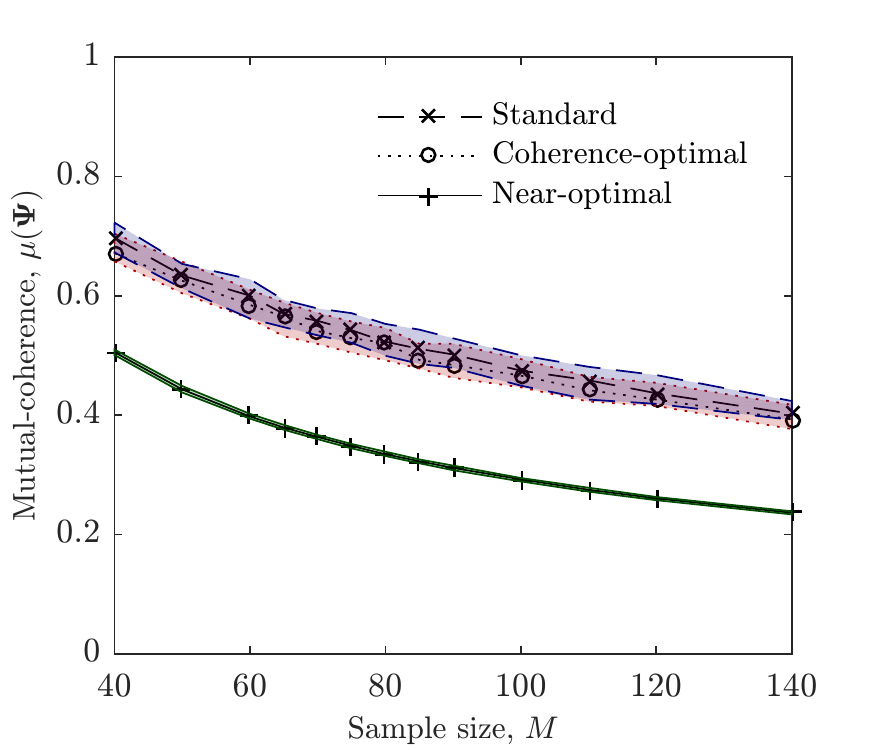}
  		\caption{}
  		\label{fig:example2-MuCoh}
  	\end{subfigure}
  	\quad
  			\begin{subfigure}[t]{0.47\linewidth}
  				\includegraphics[width=1\linewidth]{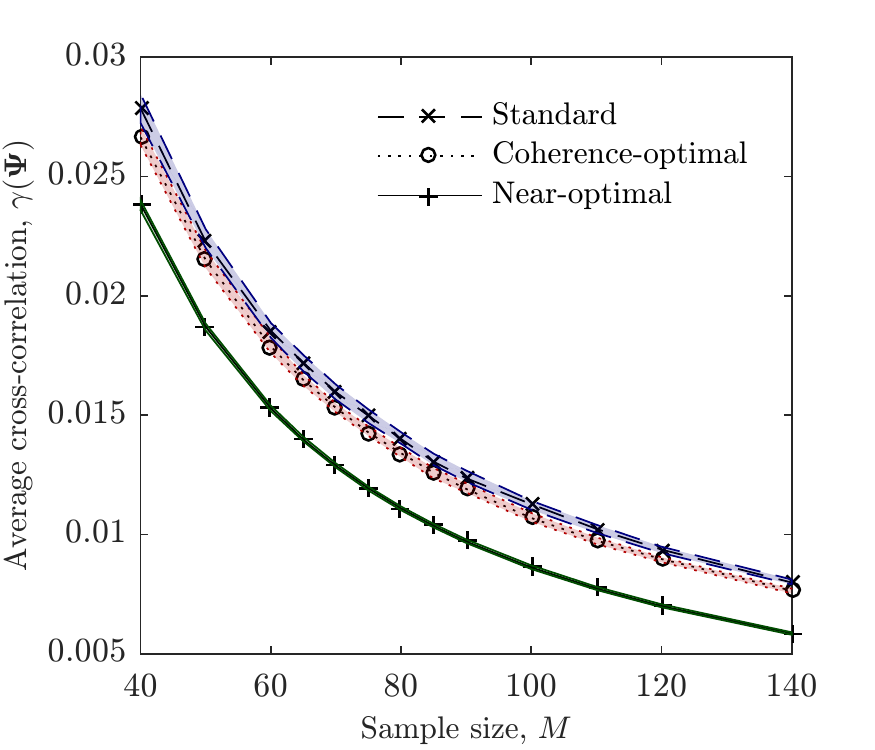}
  				\caption{}
  				\label{fig:example2-OrthMes}
  	\end{subfigure}
 
		\begin{subfigure}[t]{0.47\linewidth}
			\includegraphics[width=1\linewidth]{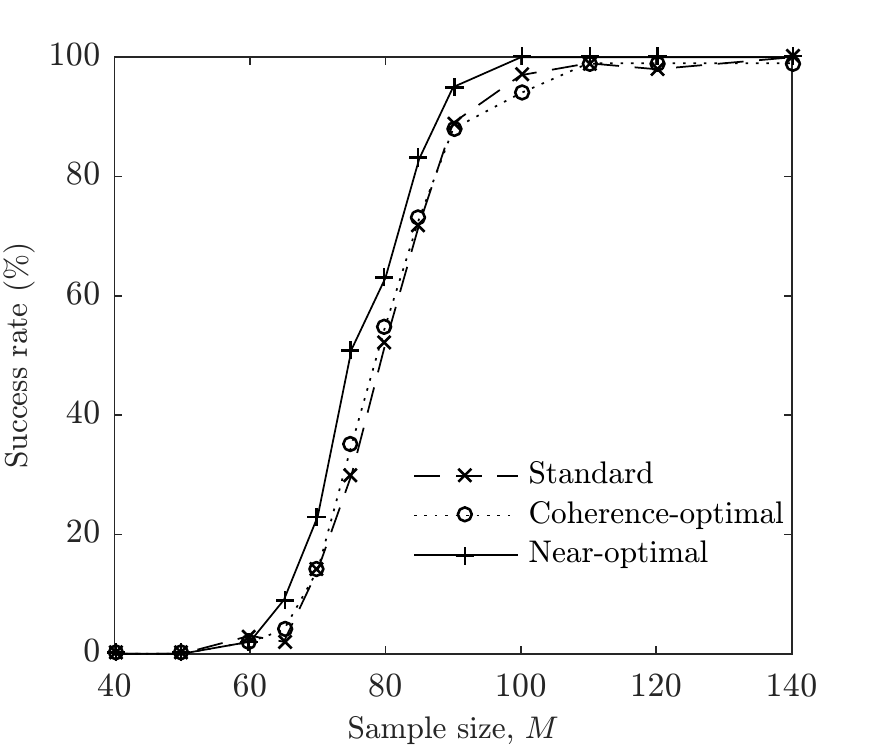}
			\caption{}
			\label{fig:example2-success}
		\end{subfigure}
 \centering
\caption{Comparison of near-optimal sampling results with  standard and coherence-optimal sampling methods for the high-dimensional low-degree manufactured PCE of Equation (\ref{ex.highDlowk}). The following measures are compared: (a)  median, and 1st and 3rd quantiles of  relative $\ell_2$ error (b)  mean of relative $\ell_2$ error, (c) median,  and 1st and 3rd quantiles of mutual-coherence, (d) median, 1st and 3rd quantiles of average cross-correlation, and (e) success rate, defined as the ratio of trials that result in a relative error smaller than $10^{-7}$.}
	\label{fig.example2.coh}
\end{figure}

As a contrasting example, let us consider the target function, $u(\bm \Xi)$, to be a sparse  second order Legendre polynomial expansion in a 20-dimensional random space with uniform density on $[-1,1]^{20}$, manufactured according to 
 \begin{equation}\label{ex.highDlowk}
 u(\bm \Xi)= \sum_{i=1}^{19} \Xi_i\Xi_{i+1}.
 \end{equation}
To compare the performance of near-optimal sampling with standard  and coherence optimal sampling on this target function, the numerical results were obtained under a setting similar to that in the previous example: 100 independent runs were used for all three approaches, and 100,000  coherence-optimal samples constituted the initial sample pool in the near-optimal approach of Algorithm \ref{pseudocode}. Figure \ref{fig:example2-error-prcs} shows the median, and 1st and 3rd quantiles of relative $\ell_2$ error. Figure \ref{fig:example2-error-mean} shows the mean of relative $\ell_2$ error. Figure \ref{fig:example2-MuCoh} and \ref{fig:example2-OrthMes} show the improvement in the median, and 1st and 3rd quantiles of mutual-coherence and average cross-correlation, respectively, when the near-optimal sampling is used, which has translated into the observed improvement in recovery accuracy. 

 It should be highlighted that standard sampling is mostly suitable for  high-dimensional and low-order cases such as this example  \cite{doostan2011non}. It should also be noted that in the examples in this section and Section \ref{low-dim-high-order}, the measurement matrices have the same number of columns ($K=231$); however, as it can be seen in Figures \ref{fig:example2-MuCoh} and \ref{fig:example2-OrthMes}, the high-dimensionality of this example has resulted in smaller mutual-coherences and average cross-correlation for various sample sizes. Consequently,  we expect to observe that recently developed sampling strategies do not outperform standard sampling in these high-dimensional low-order problem cases \cite{tang2014subsampled, jakeman2016generalized}. However, as it can be seen in figures \ref{fig:example2-error-prcs} and \ref{fig:example2-error-mean}, the near-optimal sampling approach does outperform both standard and coherence optimal sampling. Although, this improvement in terms of mean or median of relative error may not seem  significant, it should be noted that this is a rather ``extreme" case, where not only the response is high-dimensional, but its order is also very low. This could be thought of as a lower bound for the improvement offered by the near-optimal sampling approach. That is,  the fact that the near-optimal sampling strategy still shows improvement, even non-significant, for such extreme cases where standard sampling is supposed to work well, can be considered as a numerical evidence that it will most likely  outperform other sampling strategies in all the other cases. 

As we already discussed in Section \ref{optimal sampling in cs}, in order to have exact recovery we need the spark of measurement matrix to be larger than two times the number of non-zero coefficients. At small sample sizes none of the sampling approaches meet this requirement. Therefore, as it can be seen in figures~\ref{fig:example2-error-prcs} and \ref{fig:example2-error-mean}, at low sample sizes, all three sampling strategies recover poor approximations. As the number of sample increases  the chance to meet this requirement also increases; hence the variability in recovery accuracy also increases.  In near-optimal sampling approach we select sample locations such that the orthogonality of measurement matrix is improved, thereby enhancing the probability of achieving a larger spark and consequently exact recovery. To further clarify this, we demonstrate the advantage of near-optimal sampling by comparing another performance measure, namely the success rate.  The success rate, here, is defined as the ratio of trials, out of the total 100 trials, that result in a relative error smaller than $10^{-7}$.  Figure \ref{fig:example2-success} shows that near-optimal sampling  consistently results in a higher success rate, and is thus expected to produce  accurate recoveries, more frequently.  	

\subsection{Generalized Rosenbrock function } \label{sec:rosenbrock}
\begin{figure} 
  	
  	\begin{subfigure}[t]{0.47 \linewidth}
  		\includegraphics[width=1\linewidth]{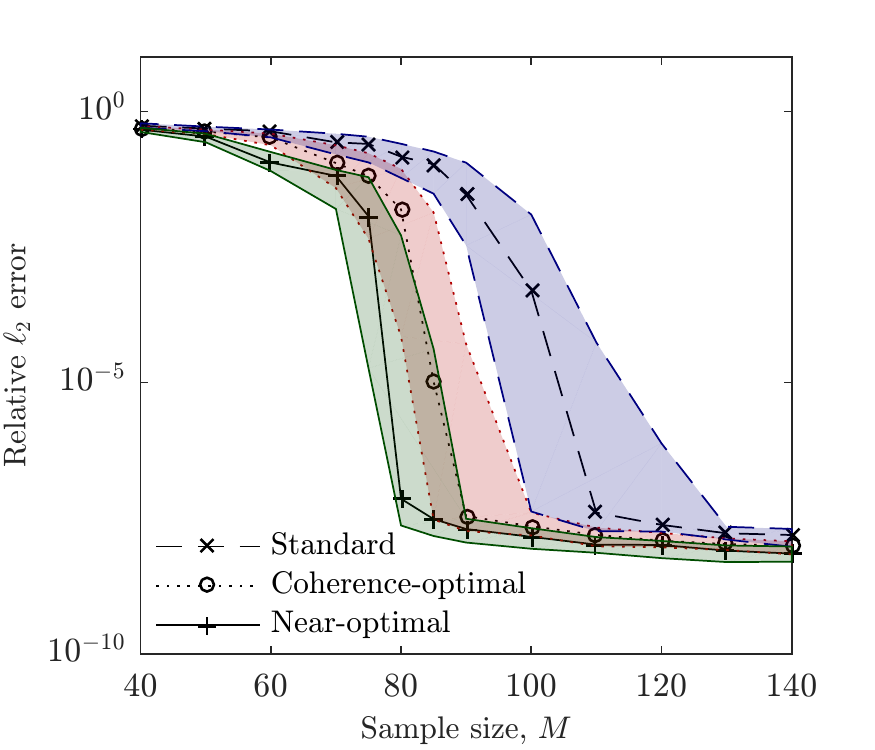}
  		\caption{}
  		\label{fig:example3-error-prcs}		
  	\end{subfigure}
  	\quad
  	\begin{subfigure}[t]{0.47 \linewidth}
  		\includegraphics[width=1\linewidth]{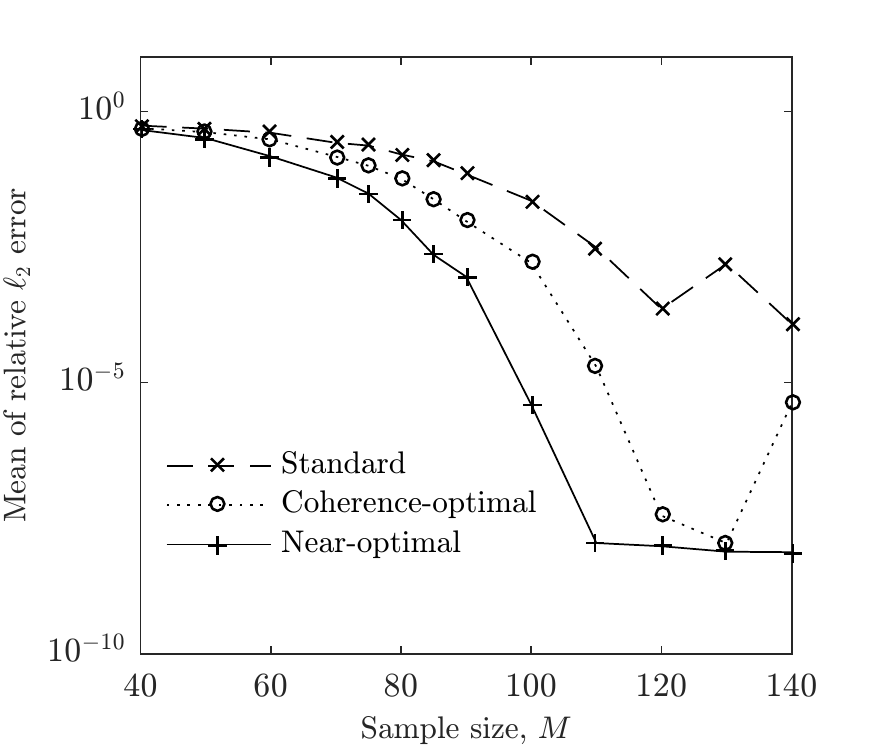}
  		\caption{}
  		\label{fig:example3-error-mean}
  	\end{subfigure}
  	
  	\begin{subfigure}[t]{0.47 \linewidth}
  		\includegraphics[width=1\linewidth]{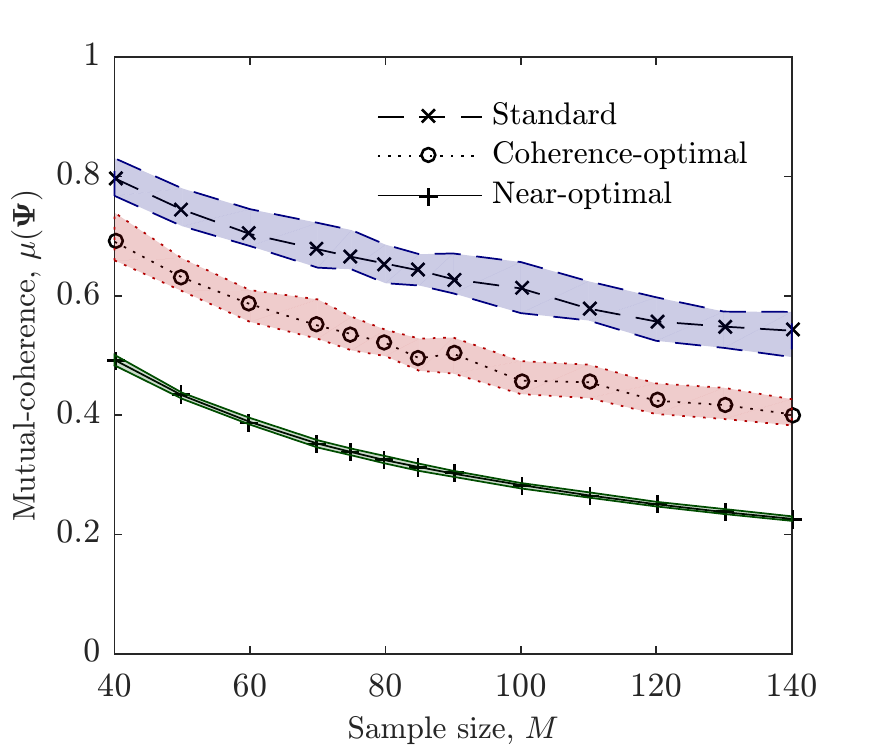}
  		\caption{}
  		\label{fig:example3-MuCoh}
  	\end{subfigure}
  	\quad
  	\begin{subfigure}[t]{0.47\linewidth}
  		\includegraphics[width=1\linewidth]{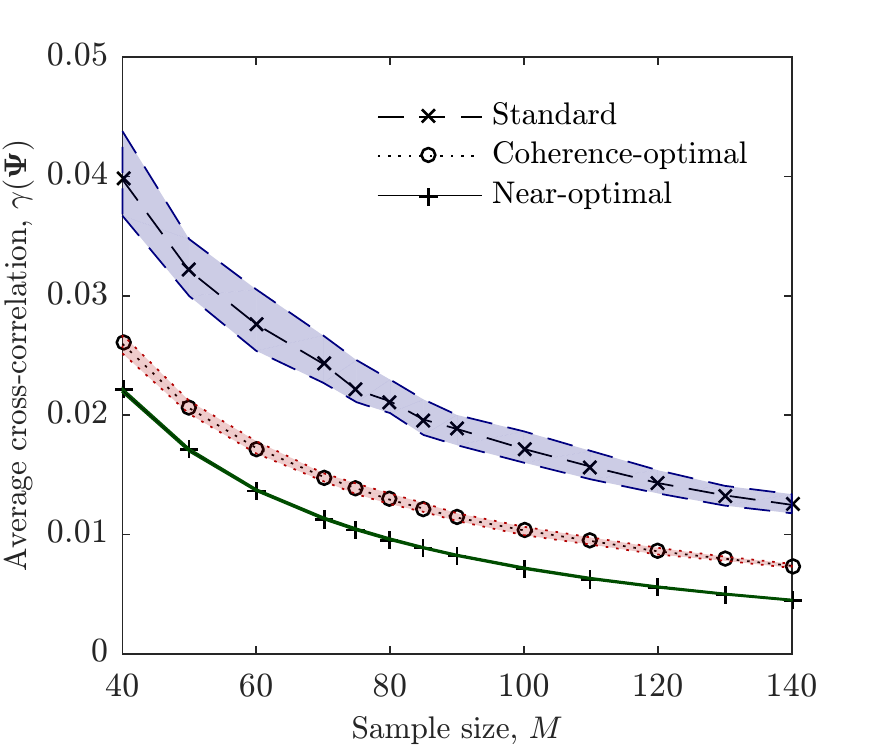}
  		\caption{}
  		\label{fig:example3-OrthMes}
  	\end{subfigure}

  		\begin{subfigure}[t]{0.47\linewidth}
  			\includegraphics[width=1\linewidth]{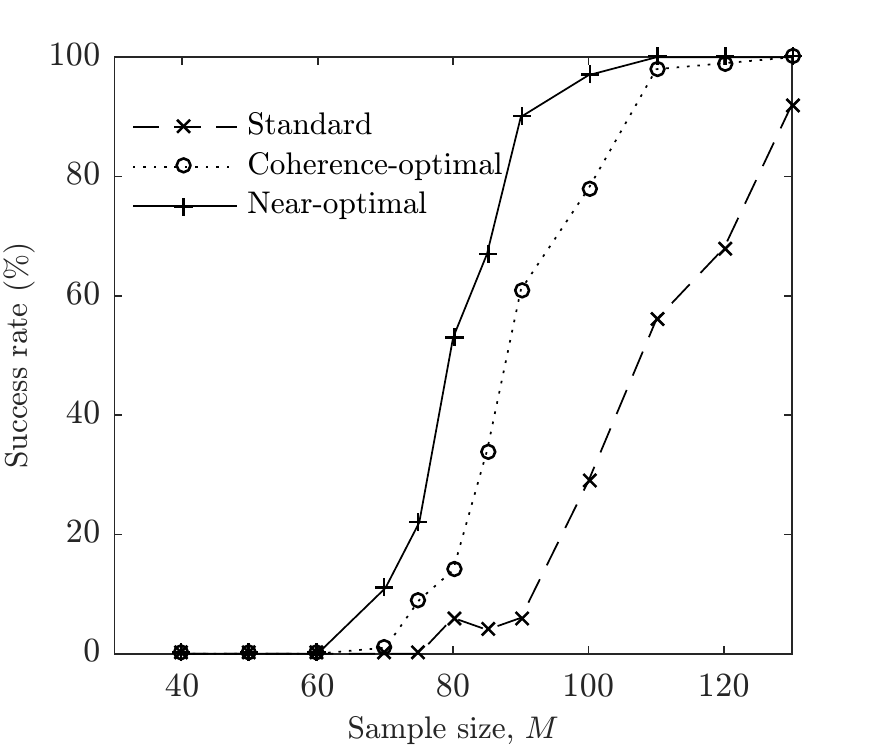}
  			\caption{}
  			\label{fig:example3-success}
  		\end{subfigure}
  		  	\centering 
\caption{Comparison of near-optimal sampling results with  standard and coherence-optimal  sampling methods for the generalized Rosenbrock function in Equation (\ref{ex.rosenbrock}). The following measures are compared: (a)  median, and 1st and 3rd quantiles of  relative $\ell_2$ error (b)  mean of relative $\ell_2$ error, (c) median,  and 1st and 3rd quantiles of mutual-coherence, (d) median, 1st and 3rd quantiles of average cross-correlation, and (e) success rate, defined as the ratio of trials that result in a relative error smaller than $10^{-7}$.}
  	\label{fig.example3}
  \end{figure}

Compared to the previous two examples with ``extreme" dimension and order combinations, the third target function can be chosen to be of ``moderate" combination of dimension and order. Specifically, let us consider the following 6-dimensional generalized Rosenbrock function with random inputs following a uniform density on $\left [ -1,1 \right ]^6$,
 \begin{equation}\label{ex.rosenbrock}
 u(\bm \Xi)= \sum_{i=1}^{5} 100(\Xi_{i+1}-\Xi_i^2)^2+(1-\Xi_i)^2.
 \end{equation}
The objective is to recover the corresponding sparse Legendre polynomial expansion.  Similar setting is used for comparison of the three sampling strategies: 100 independent runs for all sampling approaches, together with a candidate  pool of 100,000 coherence-optimal samples for near-optimal sampling of Algorithm \ref{pseudocode}. Figure \ref{fig:example3-error-prcs} demonstrates the improvements in the median, and 1st and 3rd quantiles of relative $\ell_2$ error when near-optimal sampling is used. Figure \ref{fig:example3-error-mean} demonstrates the improvements in terms of the mean of relative $\ell_2$ error. Figure \ref{fig:example3-MuCoh} and \ref{fig:example3-OrthMes} show that near-optimal sampling results in smaller mutual-coherence and average cross-correlation,  in terms of their median, and 1st and 3rd quantiles. Similar to the previous example, we note that at small sample sizes all approaches result in equally inaccurate recoveries as all approaches fail to achieve a measurement matrix with sufficiently large spark, i.e. \text{spark}$(\bm \Psi)> 2s$. A larger sample sizes results in a larger lower bound for the spark of measurement matrix and a higher probability to achieve the critical spark value. Therefore, we observe more variability in recovery accuracy as the sample size increases. Figure \ref{fig:example3-success} shows that using near-optimal sampling leads to significant improvement in the success rate, i.e. the ratio of trials with relative errors smaller than $10^{-7}$. This is a direct result of improving the orthogonality of measurement matrix, i.e. enhancing the chance of achieving critical spark value in near-optimal sampling.

\subsection{Stochastic diffusion problem}

In this example we consider a stochastic diffusion problem  in a one dimensional physical domain, given by
 \begin{equation}\label{ex.diffusion}
 \begin{aligned}
 -\frac{\partial }{\partial x} \left(a( x,\bm \Xi) \frac{\partial u}{\partial x}  ( x,\bm \Xi)\right) &=2, \quad  x \in (0,1),
 \\
 u(0,\bm \Xi) =0, \quad  u(1,\bm \Xi) &=0, \quad \bm \Xi \in [-1,1]^{10}.
 \end{aligned}
 \end{equation}
 We assume that the diffusion coefficient, $a( x,\bm \Xi)$, takes the following analytical form
 \begin{equation}
 a( x,\bm \Xi)=1+ \sum_{k=1}^{10}\frac{1}{k^2\pi^2}\cos(2\pi kx)\Xi_k.
 \end{equation}

\begin{figure}
 	
 	\begin{subfigure}[t]{0.47 \linewidth}
 		\includegraphics[width=1\linewidth]{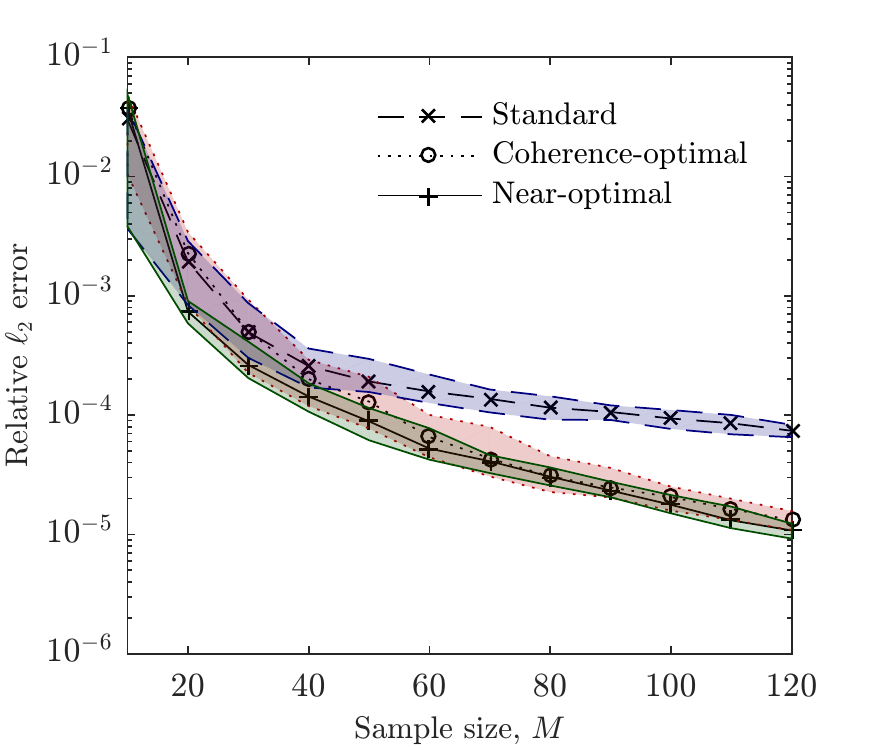}
 		\caption{}
 		\label{fig:example4-error-prcs}		
 	\end{subfigure}
 	\quad
 	\begin{subfigure}[t]{0.47 \linewidth}
 		\includegraphics[width=1\linewidth]{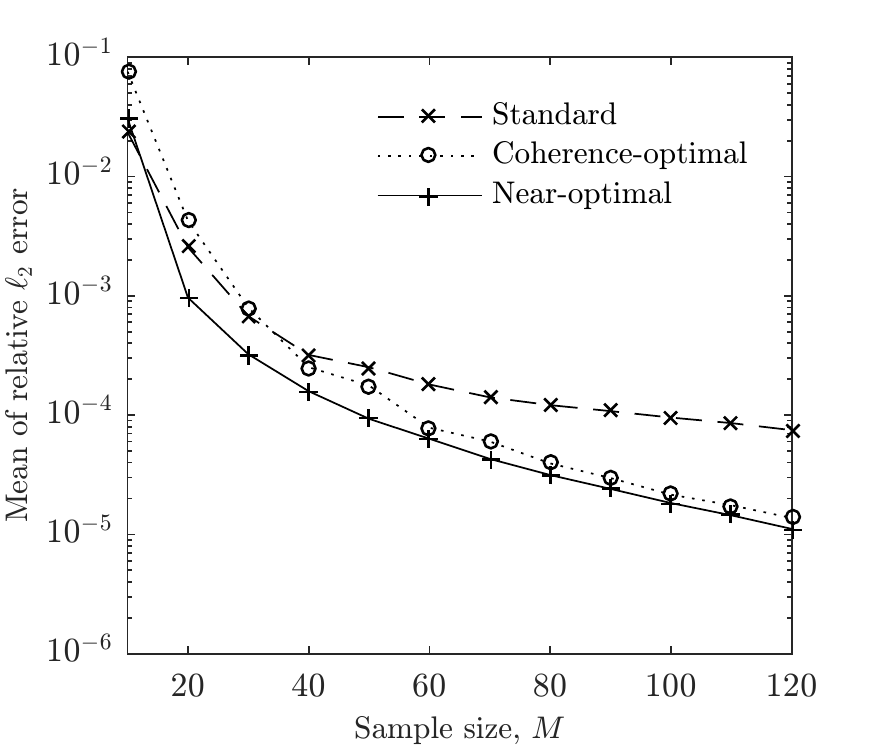}
 		\caption{}
 		\label{fig:example4-error-mean}
 	\end{subfigure}
 	
 	\begin{subfigure}[t]{0.47 \linewidth}
 		\includegraphics[width=1\linewidth]{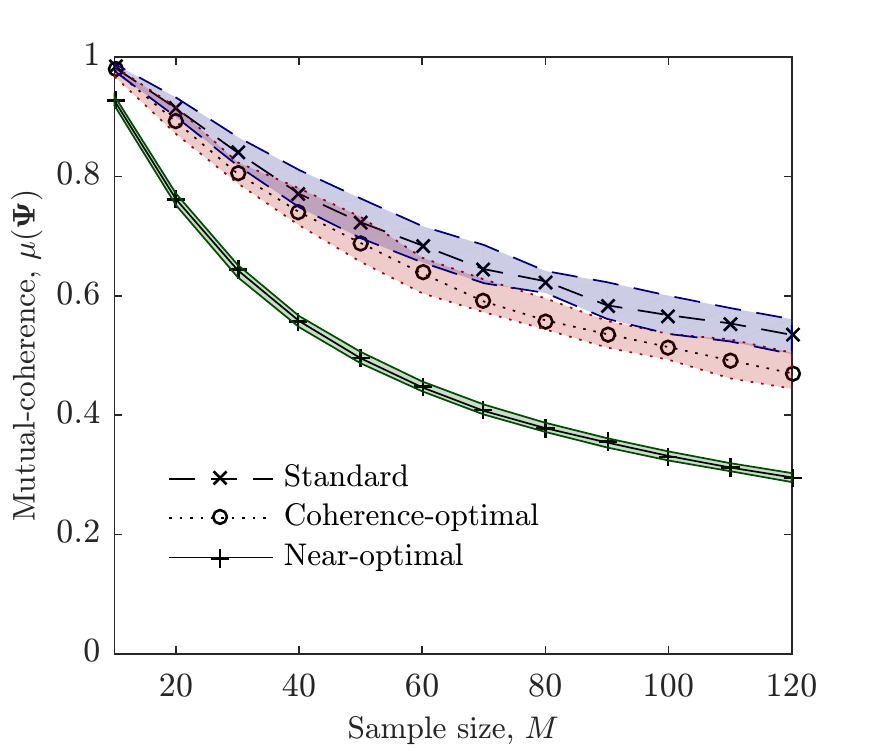}
 		\caption{}
 		\label{fig:example4-MuCoh}
 	\end{subfigure}
 	\quad
 	\begin{subfigure}[t]{0.47\linewidth}
 		\includegraphics[width=1\linewidth]{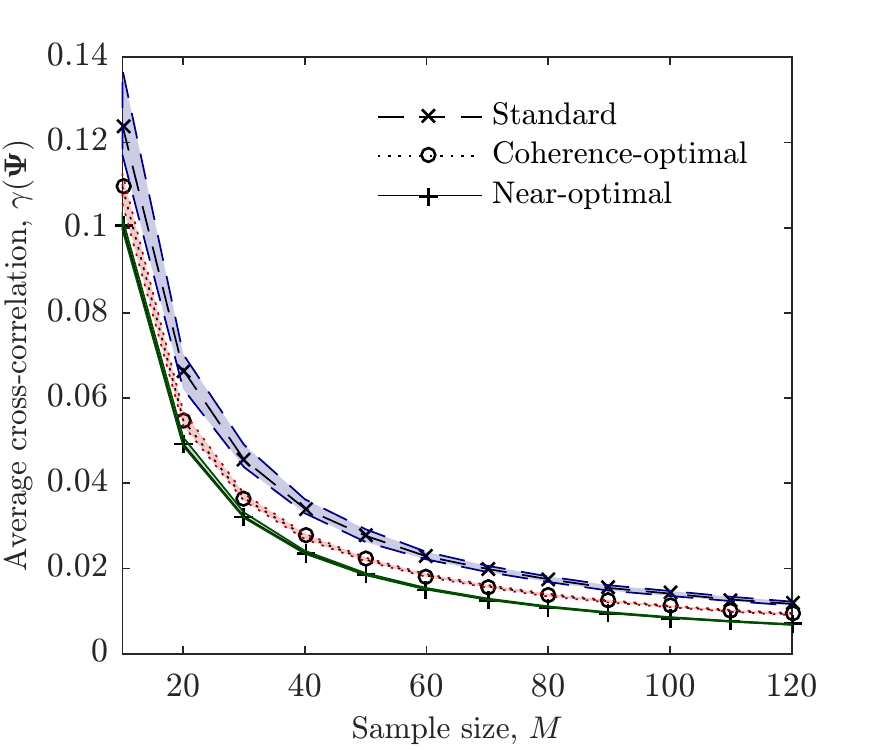}
 		\caption{}
 		\label{fig:example4-OrthMes}
 	\end{subfigure}
 	 	  
 		\begin{subfigure}[t]{0.47\linewidth}
 			\includegraphics[width=1\linewidth]{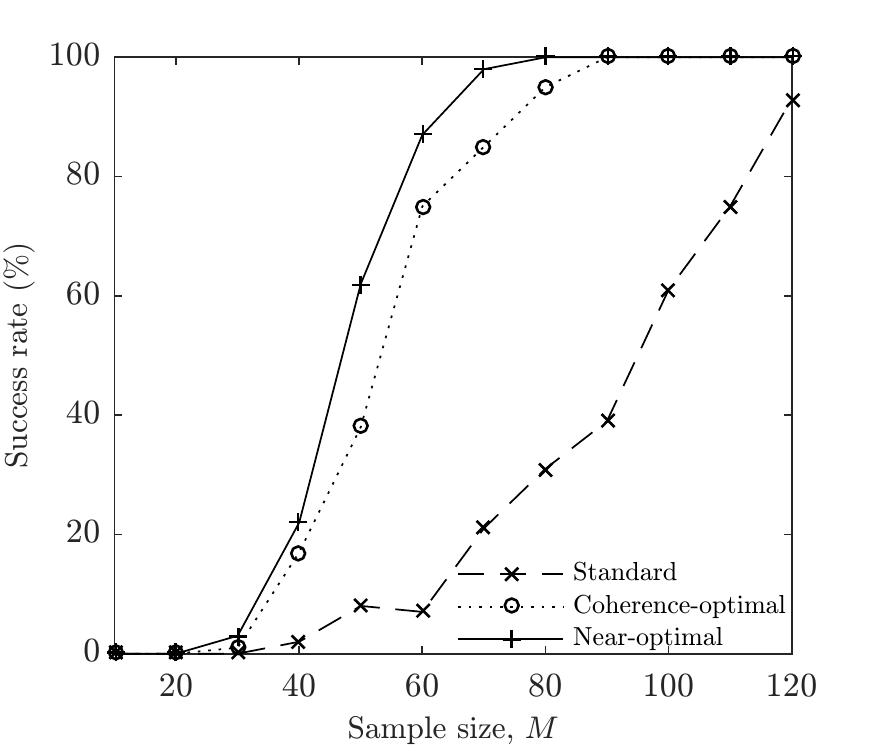}
 			\caption{}
 			\label{fig:example4-success}
 		\end{subfigure}
 	    \centering
     \caption{Comparison of near-optimal sampling results with  standard and coherence-optimal  sampling methods for the solution of the stochastic diffusion problem of Equation (\ref{ex.diffusion}). The following measures are compared: (a)  median, and 1st and 3rd quantiles of  relative $\ell_2$ error (b)  mean of relative $\ell_2$ error, (c) median,  and 1st and 3rd quantiles of mutual-coherence, (d) median, 1st and 3rd quantiles of average cross-correlation, and (e) success rate, defined as the ratio of trials that result in a relative error smaller than $10^{-4}$.}
 	\label{fig.example4}
 \end{figure}

 We consider $\bm \Xi$ to be uniformly distributed on $[-1,1]^{10}$ and consider the solution of diffusion problem at $u(0.5,\bm \Xi)$ to be the quantity of interest. We use 3rd-order Legendre polynomial expansion to approximate the QoI and employ the three sampling approaches to estimate the coefficients of expansion. Similar setting to previous examples is used here: 100 independent runs for all sampling approaches, together with a candidate  pool of 100,000 coherence-optimal samples for near-optimal sampling of Algorithm \ref{pseudocode}. For this example, we define the relative error to be $(\left \| \bm u - \bar{\bm u} \right \|_2)/(\left \| \bar{\bm u}  \right \|_2)$, where $\bar{\bm u}$ is the vector including the exact solutions calculated at 1000 new random samples and $\bm u$ includes the approximate solutions calculated by evaluating the PC expansion at the same 1000 samples.  Figure \ref{fig:example4-error-prcs} demonstrates the improvement in the median, and 1st and 3rd quantiles of relative $\ell_2$ error when near-optimal sampling is used. Figure \ref{fig:example4-error-mean} shows the improvement in the mean of relative $\ell_2$ error. The  mutual-coherence and average cross-correlation are also compared in figures \ref{fig:example4-MuCoh} and \ref{fig:example4-OrthMes}, respectively, in terms of their medians, and 1st and 3rd quantiles. To compare the success rates, since the  target expansion is not exactly sparse, we consider a looser definition for successful recovery. This is justified by noting that the error levels in Figure \ref{fig:example4-error-prcs} are relatively larger than those those in  Figures \ref{fig:example2-error-prcs} and  \ref{fig:example3-error-prcs}. Accordingly, we define a recovery to be successful when its relative $\ell_2$ error is smaller that $10^{-4}$, and show the resulting success rates   in  Figure \ref{fig:example4-success}. These results show that success rates can be  improved significantly by using near-optimal sampling.

\section{Conclusion}
In this paper, we presented a new sampling strategy, or a design-of-experiment technique, for selecting sample locations for sparse estimation of PCEs using compressive sampling. The sample locations are selected such that (i) the local-coherence property is improved, and (ii) the resulting measurement matrix has the smallest mutual-coherence and the smallest average cross-correlation between its columns. It was discussed how the latter two measures have implications on recovery accuracy and numerical results were presented to support it. A greedy algorithm was introduced that selects a prescribed number of near-optimal locations out of a large pool of candidate locations. The resulting measurement matrix is claimed to be near-optimal, rather than optimal, because  the two aforementioned cross-correlation measures may not be the only measures that control the recovery accuracy, and therefore minimizing them  does not guarantee optimal performance. Another reason why our algorithm can be sub-optimal is that the greedy algorithm is inherently a non-deterministic algorithm, and as such is susceptible to variabilities induced by random choice of location candidates and  initial (first)   sample location in the algorithm. Four numerical examples with various combinations of dimensionality and order demonstrated the advantages of our proposed sampling strategy over other sampling strategies, in terms of  accuracy and robustness.

\section{References}
\bibliography{bibfile}

\begin{thebibliography}{10}
\expandafter\ifx\csname url\endcsname\relax
  \def\url#1{\texttt{#1}}\fi
\expandafter\ifx\csname urlprefix\endcsname\relax\def\urlprefix{URL }\fi
\expandafter\ifx\csname href\endcsname\relax
  \def\href#1#2{#2} \def\path#1{#1}\fi

\bibitem{ghanem2003stochastic}
R.~G. Ghanem, P.~D. Spanos, Stochastic finite elements: a spectral approach,
  Courier Corporation, 2003.

\bibitem{xiu2002wiener}
D.~Xiu, G.~E. Karniadakis, The {Wiener--Askey} polynomial chaos for stochastic
  differential equations, SIAM journal on scientific computing 24~(2) (2002)
  619--644.

\bibitem{deb2001solution}
M.~K. Deb, I.~M. Babu{\v{s}}ka, J.~T. Oden, Solution of stochastic partial
  differential equations using {Galerkin finite} element techniques, Computer
  Methods in Applied Mechanics and Engineering 190~(48) (2001) 6359--6372.

\bibitem{babuska2004galerkin}
I.~Babuska, R.~Tempone, G.~E. Zouraris, Galerkin finite element approximations
  of stochastic elliptic partial differential equations, SIAM Journal on
  Numerical Analysis 42~(2) (2004) 800--825.

\bibitem{eldred2008evaluation}
M.~S. Eldred, C.~G. Webster, P.~Constantine, Evaluation of non-intrusive
  approaches for {Wiener-Askey} generalized polynomial chaos, in: Proceedings
  of the 10th AIAA Non-Deterministic Approaches Conference, number
  AIAA-2008-1892, Schaumburg, IL, Vol. 117, 2008, p. 189.

\bibitem{novak1997curse}
E.~Novak, K.~Ritter, The curse of dimension and a universal method for
  numerical integration, in: Multivariate approximation and splines, Springer,
  1997, pp. 177--187.

\bibitem{ganapathysubramanian2007sparse}
B.~Ganapathysubramanian, N.~Zabaras, Sparse grid collocation schemes for
  stochastic natural convection problems, Journal of Computational Physics
  225~(1) (2007) 652--685.

\bibitem{xiu2005high}
D.~Xiu, J.~S. Hesthaven, High-order collocation methods for differential
  equations with random inputs, SIAM Journal on Scientific Computing 27~(3)
  (2005) 1118--1139.

\bibitem{nobile2008sparse}
F.~Nobile, R.~Tempone, C.~G. Webster, A sparse grid stochastic collocation
  method for partial differential equations with random input data, SIAM
  Journal on Numerical Analysis 46~(5) (2008) 2309--2345.

\bibitem{candes2008introduction}
E.~J. Cand{\`e}s, M.~B. Wakin, An introduction to compressive sampling, IEEE
  signal processing magazine 25~(2) (2008) 21--30.

\bibitem{donoho2006compressed}
D.~L. Donoho, Compressed sensing, Information Theory, IEEE Transactions on
  52~(4) (2006) 1289--1306.

\bibitem{candes2006robust}
E.~J. Cand{\`e}s, J.~Romberg, T.~Tao, Robust uncertainty principles: Exact
  signal reconstruction from highly incomplete frequency information,
  Information Theory, IEEE Transactions on 52~(2) (2006) 489--509.

\bibitem{doostan2011non}
A.~Doostan, H.~Owhadi, A non-adapted sparse approximation of {PDEs} with
  stochastic inputs, Journal of Computational Physics 230~(8) (2011)
  3015--3034.

\bibitem{mathelin2012compressed}
L.~Mathelin, K.~Gallivan, A compressed sensing approach for partial
  differential equations with random input data, Communications in
  computational physics 12~(04) (2012) 919--954.

\bibitem{blatman2011adaptive}
G.~Blatman, B.~Sudret, Adaptive sparse polynomial chaos expansion based on
  least angle regression, Journal of Computational Physics 230~(6) (2011)
  2345--2367.

\bibitem{yang2016enhancing}
X.~Yang, H.~Lei, N.~A. Baker, G.~Lin, Enhancing sparsity of {Hermite}
  polynomial expansions by iterative rotations, Journal of Computational
  Physics 307 (2016) 94--109.

\bibitem{jakeman2015enhancing}
J.~D. Jakeman, M.~S. Eldred, K.~Sargsyan, Enhancing {$\ell_{1}$} -minimization
  estimates of polynomial chaos expansions using basis selection, Journal of
  Computational Physics 289 (2015) 18--34.

\bibitem{alemazkoor2016divide}
N.~Alemazkoor, H.~Meidani, Divide and conquer: an incremental sparsity
  promoting compressive sampling approach for polynomial chaos expansions,
  arXiv preprint arXiv:1606.06611.

\bibitem{rauhut2012sparse}
H.~Rauhut, R.~Ward, Sparse {Legendre} expansions via {$\ell_1$}-minimization,
  Journal of approximation theory 164~(5) (2012) 517--533.

\bibitem{tang2014subsampled}
G.~Tang, G.~Iaccarino, Subsampled {Gauss} quadrature nodes for estimating
  polynomial chaos expansions, SIAM/ASA Journal on Uncertainty Quantification
  2~(1) (2014) 423--443.

\bibitem{hampton2015compressive}
J.~Hampton, A.~Doostan, Compressive sampling of polynomial chaos expansions:
  convergence analysis and sampling strategies, Journal of Computational
  Physics 280 (2015) 363--386.

\bibitem{xiu2010numerical}
D.~Xiu, Numerical methods for stochastic computations: a spectral method
  approach, Princeton University Press, 2010.

\bibitem{hegland2003adaptive}
M.~Hegland, Adaptive sparse grids, Anziam Journal 44 (2003) 335--353.

\bibitem{ma2009adaptive}
X.~Ma, N.~Zabaras, An adaptive hierarchical sparse grid collocation algorithm
  for the solution of stochastic differential equations, Journal of
  Computational Physics 228~(8) (2009) 3084--3113.

\bibitem{shin2016nonadaptive}
Y.~Shin, D.~Xiu, Nonadaptive quasi-optimal points selection for least squares
  linear regression, SIAM Journal on Scientific Computing 38~(1) (2016)
  A385--A411.

\bibitem{lustig2005application}
M.~Lustig, J.~M. Santos, J.-H. Lee, D.~L. Donoho, J.~M. Pauly, Application of
  compressed sensing for rapid {MR} imaging, SPARS,(Rennes, France).

\bibitem{ender2010compressive}
J.~H. Ender, On compressive sensing applied to radar, Signal Processing 90~(5)
  (2010) 1402--1414.

\bibitem{paredes2007ultra}
J.~L. Paredes, G.~R. Arce, Z.~Wang, {Ultra-wideband} compressed sensing:
  channel estimation, Selected Topics in Signal Processing, IEEE Journal of
  1~(3) (2007) 383--395.

\bibitem{gemmeke2010compressive}
J.~F. Gemmeke, H.~Van~Hamme, B.~Cranen, L.~Boves, Compressive sensing for
  missing data imputation in noise robust speech recognition, Selected Topics
  in Signal Processing, IEEE Journal of 4~(2) (2010) 272--287.

\bibitem{yan2012stochastic}
L.~Yan, L.~Guo, D.~Xiu, Stochastic collocation algorithms using {$\ell_{1}$}
  minimization, International Journal for Uncertainty Quantification 2~(3).

\bibitem{yang2013reweighted}
X.~Yang, G.~E. Karniadakis, Reweighted {$\ell_{1}$} minimization method for
  stochastic elliptic differential equations, Journal of Computational Physics
  248 (2013) 87--108.

\bibitem{bruckstein2009sparse}
A.~M. Bruckstein, D.~L. Donoho, M.~Elad, From sparse solutions of systems of
  equations to sparse modeling of signals and images, SIAM review 51~(1) (2009)
  34--81.

\bibitem{chen2001atomic}
S.~S. Chen, D.~L. Donoho, M.~A. Saunders, Atomic decomposition by basis
  pursuit, SIAM review 43~(1) (2001) 129--159.

\bibitem{candes2008restricted}
E.~J. Cand{\`e}s, The restricted isometry property and its implications for
  compressed sensing, Comptes Rendus Mathematique 346~(9) (2008) 589--592.

\bibitem{maleki2010approximate}
M.~A. Maleki, D.~L. Donoho, R.~Gray, A.~Montanari, Approximate message passing
  algorithms for compressed sensing, Stanford University, 2010.

\bibitem{rauhut2010compressive}
H.~Rauhut, Compressive sensing and structured random matrices, Theoretical
  foundations and numerical methods for sparse recovery 9 (2010) 1--92.

\bibitem{davenport2011introduction}
M.~A. Davenport, M.~F. Duarte, Y.~C. Eldar, G.~Kutyniok, Introduction to
  compressed sensing, Preprint 93~(1) (2011) 2.

\bibitem{donoho2003optimally}
D.~L. Donoho, M.~Elad, Optimally sparse representation in general
  (nonorthogonal) dictionaries via {$\ell_1$} minimization, Proceedings of the
  National Academy of Sciences 100~(5) (2003) 2197--2202.

\bibitem{li2013projection}
G.~Li, Z.~Zhu, D.~Yang, L.~Chang, H.~Bai, On projection matrix optimization for
  compressive sensing systems, IEEE Transactions on Signal Processing 61~(11)
  (2013) 2887--2898.

\bibitem{cheng2015fundamentals}
H.~Cheng, The fundamentals of compressed sensing, in: Sparse Representation,
  Modeling and Learning in Visual Recognition, Springer, 2015, pp. 21--53.

\bibitem{donoho2006stable}
D.~L. Donoho, M.~Elad, V.~N. Temlyakov, Stable recovery of sparse overcomplete
  representations in the presence of noise, Information Theory, IEEE
  Transactions on 52~(1) (2006) 6--18.

\bibitem{elad2010sparse}
M.~Elad, Sparse and redundant representations: From theory to applications in
  signal and image processing.

\bibitem{elad2007optimized}
M.~Elad, Optimized projections for compressed sensing, IEEE Transactions on
  Signal Processing 55~(12) (2007) 5695--5702.

\bibitem{duarte2008learning}
J.~M. Duarte-Carvajalino, G.~Sapiro, Learning to sense sparse signals:
  Simultaneous sensing matrix and sparsifying dictionary optimization, Tech.
  rep., DTIC Document (2008).

\bibitem{abolghasemi2010optimization}
V.~Abolghasemi, S.~Ferdowsi, B.~Makkiabadi, S.~Sanei, On optimization of the
  measurement matrix for compressive sensing, in: Signal Processing Conference,
  2010 18th European, IEEE, 2010, pp. 427--431.

\bibitem{zelnik2011sensing}
L.~Zelnik-Manor, K.~Rosenblum, Y.~C. Eldar, Sensing matrix optimization for
  block-sparse decoding, IEEE Transactions on Signal Processing 59~(9) (2011)
  4300--4312.

\bibitem{tian2016orthogonal}
S.~Tian, X.~Fan, Z.~Li, T.~Pan, Y.~Choi, H.~Sekiya, Orthogonal-gradient
  measurement matrix construction algorithm, Chinese Journal of Electronics
  25~(1) (2016) 81--87.

\bibitem{zavala2012stability}
V.~M. Zavala, A.~Flores-Tlacuahuac, Stability of multiobjective predictive
  control: A utopia-tracking approach, Automatica 48~(10) (2012) 2627--2632.

\bibitem{roman2006evenly}
C.~Roman, W.~Rosehart, Evenly distributed pareto points in multi-objective
  optimal power flow, IEEE Transactions on Power Systems 21~(2) (2006)
  1011--1012.

\bibitem{van2007spgl1}
E.~Van Den~Berg, M.~Friedlander, {SPGL1}: A solver for large-scale sparse
  reconstruction (2007).

\bibitem{jakeman2016generalized}
J.~D. Jakeman, A.~Narayan, T.~Zhou, A generalized sampling and preconditioning
  scheme for sparse approximation of polynomial chaos expansions, arXiv
  preprint arXiv:1602.06879.

\end{thebibliography}

\end{document}